\documentclass[10pt,
superscriptaddress,
preprintnumbers,
nofootinbib,
amsmath,amssymb,
aps,
prd,
floatfix
]{revtex4-2}

\usepackage[pdfpagelabels, pdfencoding=auto, psdextra]{hyperref}
\hypersetup{%
 pdfsubject=Paper,
 pdfkeywords={nuclear physics} 
 unicode = true,
 breaklinks = true,
 colorlinks = true,
 linkcolor = blue,
 menucolor = blue,
 citecolor = blue,
 urlcolor = blue
}

\usepackage{float}
\usepackage{scalerel}
\usepackage{subfig}
\usepackage{graphicx}
\usepackage{dcolumn}
\usepackage{bm}
\usepackage{slashed}
\usepackage[usenames]{xcolor}
\usepackage{appendix}
\usepackage{braket}
\newcolumntype{P}[1]{>{\centering\arraybackslash}p{#1}}

\newcommand{\Li}{\text{Li}}
\newcommand{\orcid}[1]{\href{https://orcid.org/#1}{\includegraphics[scale=0.055]{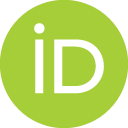}}}

\begin{document}

\title{Effects of Landau quantization on neutrino emission and absorption}
\date{\today}
\author{Mia Kumamoto~\orcid{0009-0004-9515-9213}}
\email{mialk@uw.edu}
\affiliation{Institute for Nuclear Theory, University of Washington, Seattle, WA USA}
\affiliation{Department of Physics, University of Washington, Seattle, WA USA}
\author{Catherine Welch~\orcid{0000-0002-1738-0463}}
\email{clwelch@uw.edu}
\affiliation{Department of Physics, University of Washington, Seattle, WA USA}

\begin{abstract}
    Some neutron stars known as magnetars possess very strong magnetic fields, with surface fields as large as $10^{15}\, \mbox{G}$ and internal fields that are possibly stronger~\cite{Olausen_2014_magnetar_catalogue}. Recent observations of the radio pulsar GLEAM-X J1627 suggest it may have a surface field as strong as $10^{16} \, \mbox{G}$~\cite{Suvorov_2023_GLEAM_magnetar}. In the presence of a strong magnetic field, the energy levels of electrons and protons are quantized and the Direct Urca process allows neutron stars to cool rapidly, even at low density~\cite{Leinson_perez,Baiko_Yakovlev_1999,yakovlev_kaminker_cooling_formulas}. For the case of magnetic fields $B \gtrsim 10^{16}\, \mbox{G}$, we find features in the emissivity due to energy quantization that are not captured by the frequently employed quasiclassical approximation where energy levels are treated as nearly continuous. Resonances can result in amplification of the neutrino emissivity at specific densities compared to a calculation that neglects quantization, particularly at low temperature. These effects are not important for the thermal evolution of an entire neutron star, but may be relevant for phenomena that depend on behavior at specific densities. We present a fully relativistic calculation of the Direct Urca rate in a strong magnetic field using the standard V-A weak Lagrangian incorporating mean field nuclear effects and discuss approaches to the numerical challenge the modified wavefunctions present and a new semi-analytic approximation. These tools are also applicable to calculating neutrino opacities in strong magnetic fields in the ejecta of binary neutron star mergers. We calculate the opacities for neutrinos capturing on free nucleons at sub-saturation densities and temperatures exceeding an MeV. We find an enhancement to capture processes of the lowest energy neutrinos by an order of magnitude or more due to suppression of electron Pauli blocking in the case of capture on neutrons, and from the effect of the nucleon magnetic moments in the case of capture on protons.
\end{abstract}
\maketitle

\section{Introduction}
\label{sec:introduction}
Young neutron stars (NS) cool primarily via emission of neutrinos produced by weak reactions in nuclear matter (see Refs.~\cite{yakovlev_kaminker_cooling_formulas, Page_2004, PAGE2006497} for reviews).  One of the fastest cooling channels available is the Direct Urca (DU) process, consisting of the reactions $n \rightarrow p + e + \overline{\nu}$ and $p+e \rightarrow n + \nu$.  This mechanism is very efficient but can only operate at high density because the Fermi momenta of the particles must obey the so-called triangle inequality ($|k_{Fn}| < |k_{Fp}| + |k_{Fe}|$) in order to conserve momentum. Due to constraints of $\beta$ equilibrium and charge neutrality in NS matter, this is only possible for large proton fraction ($Y_p > 1/9$ for matter containing neutrons, protons, and electrons). At sufficiently high density (the ``DU threshold"), DU becomes available causing more massive NSs to cool rapidly. Below this threshold, slower neutrino emission processes dominate.  Modified Urca (MU), in which a spectator nucleon conserves momentum, and neutral current Bremsstrahlung are the primary sources of neutrino cooling at these densities \cite{Page_2004}.

NSs are known to possess strong magnetic fields, with typical NSs having a surface field of $10^{12-13}\, \mbox{G}$ or less with a subset of NSs known as magnetars possessing surface fields of up to $10^{15}\, \mbox{G}$~\cite{Enoto_2019, Manchester_2005_pulsar_catalogue, Olausen_2014_magnetar_catalogue}. Recent observations of the radio pulsar GLEAM-X J1627 suggest it may have a surface field as strong as $10^{16} \, \mbox{G}$, motivating the study of neutrino processes at such strong magnetic fields~\cite{Suvorov_2023_GLEAM_magnetar}. While the internal magnetic field configuration of magnetars is not fully known, (see Ref.~\cite{Dexheimer_2017_field_configurations} for a detailed discussion) it is expected that magnetic fields inside magnetars may be several times larger than at their surface. Although the protons in the cores of NSs are expected to be superconducting, it is likely that the field of a magnetar is stronger than the critical field strength and the Meissner effect does not expel the magnetic field in the core \cite{universeMagnetar}.

In the presence of strong magnetic fields, electrons and protons have their momentum perpendicular to the field quantized into Landau levels (LL)~\cite{Canuto_chiu_1968}. The momentum-space wavefunction associated with each LL has a long tail allowing momentum to be conserved in DU reactions even at low densities \cite{Leinson_perez, Baiko_Yakovlev_1999}. For higher energy states, denoted by quantum number $n$, there is a larger contribution from high momentum components.  Higher values of $n$ correspond to larger amounts of energy stored in motion perpendicular to the field, in analogy with larger momenta perpendicular to the field for a free particle. 

When many LL are available, one can make the quasiclassical (QC) approximation and treat the levels as nearly continuous. The DU rate in the presence of a strong magnetic field has been calculated in Refs.~\cite{Leinson_perez, Baiko_Yakovlev_1999, yakovlev_kaminker_cooling_formulas} in the QC approximation, giving useful formulae at NS densities for magnetic fields $ B \lesssim 10^{15} \, \mbox{G}$, and in the limit of super-strong fields $B \gtrsim 10^{18}\, \mbox{G}$ where all of the charged particles are confined to the lowest LL. In this work, we address the region in between the quasiclassical and strong field limits where a finite number of LL are occupied and quantization effects cannot be neglected. Recent work in Ref.~\cite{tambe2024effectmagneticfieldsurca} extended the quasiclassical result to high temperatures relevant for NS mergers, incorporating a finite isospin chemical potential necessary at high temperatures first discussed in Refs.~\cite{AlfordHarrisPRC, universeAlfordHaberHarris}. Our calculation differs from theirs in that our matrix element is fully relativistic and we consider lower temperatures where the isospin chemical potential is insignificant, but where quantization effects are important and the proton anomalous magnetic moment cannot be neglected. A similar calculation was also performed in Ref.~\cite{MARUYAMA2022136813} where mean field and relativistic effects were included, taking the low temperature limit and setting the momenta of the particles to their Fermi momenta. We find that even at temperatures as low as a few keV, this is a troublesome approximation due to quantization effects and also perform our calculation higher temperatures than they considered.

The case of a finite number of populated LL has been studied in the context of supernovae~\cite{Duan_qian_2004, duan_qian_2005_appdx_has_integrals} where neutrino processes in hot, low density matter were considered, including terms up to $\mathcal{O}(k/M_N)$ for $M_N$ the nucleon mass. Those works found modest modifications to the neutrino cross section at field strengths of $10^{16} \, \mbox{G}$ and temperatures of a few MeV. This work is complementary, first considering high density matter at keV to MeV temperatures and later considering low density matter at temperatures exceeding an MeV and stronger fields. Our high density calculation is fully relativistic for the V-A weak Lagrangian going beyond $\mathcal{O}(k/M_N)$ used by previous authors, but does not include additional terms in the weak nucleon current as is done in Ref.~\cite{duan_qian_2005_appdx_has_integrals}. We defer this improvement to future work. At low density, we apply the same tools for stronger fields and at similar and higher temperatures than considered by Refs.~\cite{Duan_qian_2004, duan_qian_2005_appdx_has_integrals} which may have applications to simulations of binary NS mergers. 

In Sec.~\ref{sec:lagrangian} we describe our nuclear Lagrangian and the modifications the magnetic field and nuclear mean field potentials make to the nucleon and lepton wavefunctions. In Sec.~\ref{sec:durcarate} we calculate the full DU emissivity, including the effects of interactions, relativity, and Landau quantization (LQ). The calculation of Fermi-Dirac factors and integration over phase space is complicated by singularities in the density of states and must be treated carefully. We present a semi-analytic (SA) approximation to this calculation and discuss computational approaches to performing the full integration in each Landau level. We then give results for the DU rate in NSs. Turning to neutrino absorption, Sec.~\ref{sec:opacity} gives expressions for neutrino opacities at low density and discusses appropriate approximations for this energy regime. We then present results for the neutrino opacity, comparing to prior results that were calculated at more modest magnetic field.  Sec.~\ref{sec:conclusion} summarizes our results and concludes.

\section{Nuclear model and wavefunctions}
\label{sec:lagrangian}
To include the effects of nuclear interactions, we use a relativistic mean field model (RMF) (for a review, see Ref.~\cite{PRC_dutra_rmf_review}).  The charged particle species also couple to the electromagnetic field.
\begin{equation}
    \begin{split}
        \mathcal{L} &= \sum_{i=n,p} \overline{\psi}_i \left[ i \slashed{\partial} - g_\omega \omega \gamma^0  - \frac{1}{2} g_\rho \rho \gamma^0 \tau_3 + e \frac{1 + \tau_3}{2} \slashed{A} - (M_N - g_\sigma \sigma)\right] \psi_i + \mathcal{L}_{\sigma \omega \rho} \\
        &+ \frac{G_F \cos \theta_c}{\sqrt{2}} (L_\mu^\dagger N^\mu + N_\mu^\dagger L^\mu) + \sum_{\ell = e,\mu} \overline{\psi}_\ell (i\slashed{\partial} - e \slashed{A} - m_\ell) \psi_\ell + \overline{\psi}_\nu i \slashed{\partial} P_L \psi_\nu - \frac{1}{4} F_{\mu \nu} F^{\mu \nu} 
    \end{split}
\end{equation}
where $\tau_3$ is the third Pauli matrix in isospin space, $P_L$ projects onto the left-handed neutrino, $G_F$ is the Fermi constant, and the Cabibbo angle $\theta_c \approx 13^\circ$. We use the standard notation $\slashed{v} = v^\mu \gamma_\mu$. The charged weak currents $L^\mu$ and $N^\mu$ are given by
\begin{equation}
    L^\mu = \overline{\psi}_\nu \gamma^\mu ( 1- \gamma^5) \psi_e
\end{equation}
\begin{equation}
    N^\mu = \overline{\psi}_p \gamma^\mu (g_V - g_A \gamma^5) \psi_n
\end{equation}
where $g_V = 1$ and $g_A = 1.27$ are the vector and axial vector form factors of the nucleon. Since the contribution of muons to any weak processes is suppressed by their large mass and low density, we do not consider muonic charged current interactions. The $\sigma$, $\omega$, and $\rho$ meson fields take on their mean field value given by their equations of motion, enforcing charge neutrality and beta equilibrium. These modify the in-medium effective mass and dispersion of the nucleons given by $E(k) = \sqrt{k^2 + M^{*2}} + U$ where $M^* = M_N - g_\sigma \langle \sigma \rangle$ and $U = g_\omega \langle \omega \rangle \pm  g_\rho \langle \rho \rangle /2$ where the plus sign is for protons and the minus for neutrons. $\mathcal{L}_{\sigma \omega \rho}$ contains the free Lagrangian for the meson fields as well as meson-meson interactions which are tuned to reproduce known properties of nuclear matter and finite nuclei (see Refs.~\cite{PRC_dutra_rmf_review, AGRAWAL20121} for details). We use the IUFSU$^*$ choice of RMF parameters from Ref.~\cite{AGRAWAL20121} because it produces NSs with maximum mass $M_{\rm TOV} \simeq 2 M_\odot$ and $R_{1.4} \simeq 12 \, \mathrm{km}$, in line with observational constraints~\cite{NANOGrav:2019jur,Antoniadis:2013pzd,Demorest:2010bx,LIGOScientific:2018cki,De:2018uhw,Capano:2019eae,Miller:2019cac,Riley:2019yda}. Additionally, the Direct Urca threshold for IUFSU$^*$ is approximately $4n_{\rm sat}$, well above the central density of the canonical $1.4 M_\odot$ NS we shall focus on. For fields less than $10^{18} \, \mbox{G}$, the calculated equation of state, particle densities, and Fermi momenta are nearly identical whether or not one includes the effects of LQ. We do not implement anomalous magnetic moments at the Lagrangian level and instead modify the dispersion following Ref.~\cite{Steinmetz2019KGPvsDP}. This has the advantage of making shifts to available LL intuitive, but produces discrepancies with a more detailed microphysical approach at order $eB/M^2$ (cf. Refs.~\cite{Broderick_2000PLmageos, MARUYAMA2022136813}) which are negligible for field strengths we consider. This gives a dispersion for the proton:
\begin{equation}
    E_p(k_{zp},n_p,s) = \sqrt{k_{zp}^2 + M^{*2} + 2n_p eB - (g-2)eBs} + U_p
\end{equation}
where $s$ is the spin of the proton ($\pm 1/2$) and $g$ is the gyromagnetic ratio of the proton. For the proton, $g \simeq 5.6$, making this an important contribution when determining when new LLs become available for each choice of spin. The electron and muon have $g \simeq 2.002$ and the anomalous magnetic moment can be ignored. The neutron also has an anomalous magnetic moment of a similar scale to the proton, but since the density of states for the neutron is smooth, the effect of the anomalous magnetic moment is unimportant at high density, contributing about an MeV to the energy of the neutron at the strongest field strengths we consider. At low density, the magnetic energy of the neutron is a sizable fraction of the neutron energy and of the same order as the mass splitting of the proton and neutron and must be included.

Following Ref.~\cite{Canuto_chiu_1968} we consider a magnetic field in the $\hat{z}$ direction and a vector potential in a symmetric gauge.
\begin{equation}
    \vec{A} = \frac{B}{2} (-y, x, 0)
\end{equation}
Solving the Dirac equation with such a potential gives quantized wavefunctions for the electron and proton. 
\begin{equation}
    I_{n,r}(x) \equiv \sqrt{\frac{r!}{n!}} e^{-x/2} x^{(n-r)/2} L_r^{n-r} (x)
\end{equation}
$L_r^{n-r}$ is a generalized Laguerre polynomial, $n$ indexes the LL of the wavefunction, and $r \leq n$ indexes the degeneracy of each level. If $n$ or $r$ is negative, this is set to zero. The dimensionless argument $x$ takes the form $eBx_\perp^2/2$ in the wavefunction where $x_\perp$ is the radial coordinate perpendicular to the direction of the magentic field.  

We add modifications to the spinors to clarify the normalization and include mean field effects. Note that we retain the dimension of the space $L$ to make clear how the unusual dimensional scaling of the emissivity is resolved. From here on, we use $M$ to refer to the in-medium effective mass of the nucleon $M = M_N - g_\sigma \langle \sigma \rangle$ and $M^*_L$ to refer to the Landau effective mass $M^*_L = k / (\partial E/\partial k)$. For leptons, there are no modifications to the dispersion from interactions and the Landau effective mass is just the energy of the lepton. For nucleons in a mean field model, $M^*_L=E - U$. The electron wavefunction is given by the following~\cite{Canuto_chiu_1968}.
\begin{equation}
\psi_e = \sqrt{E_e} \frac{e^{-i(E_et-k_{ze} z)} e^{i(n_e-r_e)\phi}}{\sqrt{2\pi L/eB}}  u_e^{(s)}
\end{equation}
\begin{equation}
u_e^{(\uparrow)} = \begin{bmatrix}
e^{-i\phi} I_{n_e-1,r_e}(eB\xi^2/2) \\[6pt] 0 \\[6pt]
\scaleto{\frac{k_{ze} }{E_e}}{18pt} e^{-i\phi} I_{n_e-1,r_e}(eB\xi^2/2) \\[6pt]
\scaleto{\frac{i \sqrt{2 n_e eB}}{E_e}}{20pt} I_{n_e, r_e}
\end{bmatrix}, \,
u_e^{(\downarrow)} = \begin{bmatrix}
0 \\[6pt]
I_{n_e,r_e}(eB\xi^2/2) \\[6pt]
-\scaleto{\frac{i \sqrt{2 n_e eB}}{E_e}}{20pt} e^{-i\phi}I_{n_e-1, r_e} \\[6pt]
-\scaleto{\frac{k_{ze} }{E_e}}{18pt} I_{n_e,r_e}(eB\xi^2/2) \\
\end{bmatrix}
\end{equation} 
Note that the electron spin up spinor is zero if $n=0$ since there is no spin up state in the lowest LL. The proton wavefunctions are the same as for the electron, but with a non-zero mass, nuclear interactions, and flipped charge.
\begin{equation}
\psi_p = \sqrt{M^*_{Lp} + M} \frac{e^{-i(E_p t-k_{zp} z)} e^{-i(n_p-r_p)\phi}}{\sqrt{2\pi L/eB}}  u_p^{(s)}
\end{equation}
\begin{equation}
u_p^{(\uparrow)} = \begin{bmatrix}
I_{n_p,r_p}(eB\xi^2/2) \\[6pt] 0 \\[6pt]
\scaleto{\frac{k_{zp} }{M^*_{Lp}+M}}{22pt} I_{n_p,r_p}(eB\xi^2/2) \\[6pt]
-\scaleto{\frac{i \sqrt{2 n_p eB}}{M^*_{Lp}+M}}{24pt} e^{i \phi} I_{n_p-1, r_p}
\end{bmatrix}, \,
u_p^{(\downarrow)} = \begin{bmatrix}
0 \\[6pt]
e^{i\phi} I_{n_p-1,r_p}(eB\xi^2/2) \\[6pt]
\scaleto{\frac{i \sqrt{2 n_p eB}}{M^*_{Lp}+M}}{24pt} I_{n_p, r_p} \\[6pt]
-\scaleto{\frac{k_{zp} }{M^*_{Lp}+M}}{22pt} e^{i\phi} I_{n_p-1,r_p}(eB\xi^2/2) \\
\end{bmatrix}
\end{equation}
The neutron and neutrino spinors are standard. 
\begin{equation}
    \psi_n = \sqrt{M^*_{Ln}+M} \frac{e^{-ik_n \cdot x}}{L^{3/2}} u_n^{(s)}
\end{equation}
\begin{equation}
    u_n^{(s)} = \begin{bmatrix}
        \chi^{(s)} \\
        \scaleto{\frac{\sigma \cdot k}{M^*_{Ln}+M}}{20pt} \chi^{(s)}
    \end{bmatrix}
\end{equation}
where $\chi^{(s)}$ is a two component vector in spin space. The antineutrino wavefunction is similar.
\begin{equation}
    \psi_{\overline{\nu}} = \sqrt{E_\nu} \frac{e^{ik_\nu \cdot x}}{L^{3/2}} v_\nu^{(s)}
\end{equation}
\begin{equation}
    v_\nu^{(s)} = \begin{bmatrix}
        -\scaleto{\frac{\sigma \cdot k_\nu}{E_\nu}}{20pt} \chi^{(s)} \\
        \chi^{(s)}
    \end{bmatrix}
\end{equation}
The positron and neutrino wavefunctions are analogous. We will use the following identities along with standard trace techniques to calculate the matrix element:
\begin{equation}
    \sum_s u_n^{(s)} \overline{u}_n^{(s)} = \frac{\tilde{\slashed{k}}_n + M}{M^*_{Ln}+M}
\end{equation}
\begin{equation}
    \sum_x v_\nu^{(s)} \overline{v}_n^{(s)} = \frac{\slashed{k}_\nu}{E_\nu}
\end{equation}
The tilde indicates additional mean field effects on the normal spin sum.
\begin{equation}
    \tilde{k}_n = \begin{bmatrix}
        M^*_{Ln} & k_{xn} & k_{yn} & k_{zn}
    \end{bmatrix}
\end{equation}

\section{Direct Urca Emissivity}
\label{sec:durcarate}

\subsection{Reduced matrix element}
To summarize the contributions of the wavefunctions and spatial integrations, we calculate a reduced matrix element for $\beta$ decay defined as
\begin{equation}
\label{eq:redmatelt}
    \begin{split}
        \mathcal{M}_{\rm red} &= (M^*_{Ln} +M) \frac{eB}{2\pi L^2} \sum_{spins} \sum_{r_e, r_p} \Big| \int d^2x_{\perp}e^{i (k_{\perp n}-k_{\perp \nu}) \cdot x_{\perp}}e^{-i(n_e-r_e-n_p+r_p)\phi} \\
        &\times (\bar{u}_p \gamma^{\mu} (g_V-g_A \gamma^5) u_n) (\bar{u}_e \gamma_{\mu}(1-\gamma^5)v_{\nu}) \Big|^2 \, .
    \end{split} 
\end{equation}
Since the neutrino momentum is order $T$ and the neutron momentum is order $\sqrt{2M \varepsilon_F}$ at high density and $\sqrt{2M T}$ at low density where $\varepsilon_F$ is the Fermi energy of the neutron, make the approximation $k_{\perp n} - k_{\perp \nu} \approx k_{\perp n}$.  Using the following formula from Refs.~\cite{integralsseriesproducts, duan_qian_2005_appdx_has_integrals}, the spatial integration can be carried out.
\begin{equation}
\begin{split}
&\int_0^{\infty} x_\perp d x_\perp \int_0^{2 \pi} d\phi \, e^{i k_{\perp n} \cdot x_{\perp}-i(n_e-r_e-n_p+r_p)\phi} I_{n_p,r_p}(eBx_\perp ^2/2) I_{n_e,r_e}(eBx_\perp ^2 /2)\\
&=\frac{2\pi}{eB} i^{n_e-r_e-n_p+r_p} e^{-i(n_e-r_e-n_p+r_p)\phi_n} I_{n_e,n_p}(k^2_{\perp n}/2eB)I_{r_e,r_p}(k^2_{\perp n }/2eB)
\end{split}
\end{equation}
This allows us to make many simplifications.  Note that while some elements in the spinor may have index $n$ or $n-1$, all of them have the same $r$, so every term in $\mathcal{M}_{\rm red}$ has a common factor $I_{r_e, r_p}^2$.  Using the following two identities, this term can be simplified:
\begin{equation}
    \sum_r I_{n,r}(x) I_{n',r}(x) = \delta_{n,n'}
\end{equation}
\begin{equation}
    \sum_r 1 = \frac{eBL^2}{2\pi}
\end{equation}
The normalization $eB/2\pi L^2$ exactly cancels with this term and the prefactor from the spatial integration.  Additionally, notice that all final terms in the sum will have the same factors of $i$ and $e^{i\phi_n}$ except for the terms with explicit $e^{\pm i\phi} I_{n-1,r}$.  After cancelling out the global phase, the matrix element can be calculated simply by taking the spin sum of the product of the currents in the second line of Eq.~\eqref{eq:redmatelt} squared, with the following substitutions.
\begin{equation}
    e^{\pm i\phi} \rightarrow \mp i e^{\pm i \phi_n}, \,
    I_{n,r} I_{n',r'} \rightarrow I_{n,n^\prime}
\end{equation}
To analytically continue to cases where $n-n'<0$, use the following identity.
\begin{equation}
    I_{n,n^\prime}(x) = (-1)^{n-n^\prime} I_{n^\prime,n}
\end{equation}
We use a trick to calculate the contribution of the spinors for the charged particles. Define a set of four-vectors $a$, $b$, $c$, and $d$:
\begin{equation}
    \sum_{s} u_e^{(s)} \overline{u}_e^{(s)} = \slashed{a}_e + \gamma^0 \slashed{b}_e + \slashed{c}_e \gamma^5 + \gamma^0 \slashed{d}_e \gamma^5
\end{equation}
\begin{equation}
    u_p^{(s)} \overline{u}_p^{(s)} = \slashed{a}_p^{(s)} + \gamma^0 \slashed{b}_p^{(s)} + \slashed{c}_p^{(s)} \gamma^5 + \gamma^0 \slashed{d}_p^{(s)} \gamma^5
\end{equation}
While it is convenient to sum over spins for the electron, for the proton the matrix element for each spin must be calculated separately since the large anomalous magnetic moment of the proton means different LL are available for each spin. This decomposition of the matrix is equivalent to a decomposition into scalar, vector, tensor, axial vector, and pseudoscalar terms, but is written in such a way to easily utilize trace identities. The spin sums for charged particles in a magnetic field has been calculated in Ref.~\cite{bhattacharya_thesis} (see also the calculation in Ref.~\cite{MARUYAMA2022136813}), but we prefer this approach as it makes the traces more straightforward and the intermediate expressions more compact. The coefficients of $a-d$ are easily found with the help of Mathematica or equivalent.

In terms of these vectors, the square of the currents can be found simply. The nucleon current gives the following.
\begin{equation}
    \begin{split}
        (\overline{u}_p^{(s)} \gamma^\alpha (g_V-g_A &\gamma^5) u_n)(\overline{u}_p^{(s)} \gamma^\beta (g_V-g_A \gamma^5) u_n)^\dagger = (g_V^2+g_A^2) \mbox{Tr}[\gamma^\alpha \tilde{\slashed{k}}_n \gamma^\beta (\slashed{a}_p^{(s)} + \slashed{c}_p^{(s)} \gamma^5) ] \\
        &+ 2 g_V g_A \mbox{Tr} [ \gamma^\alpha \tilde{\slashed{k}}_n \gamma^\beta (\slashed{c}_p^{(s)} + \slashed{a}_p^{(s)} \gamma^5)] + M (g_V^2-g_A^2) \mbox{Tr}[ \gamma^\alpha \gamma^\beta \gamma^0 (\slashed{b}_p^{(s)} + \slashed{d}_p^{(s)} \gamma^5)]
    \end{split}
\end{equation}
The leptonic current is given by
\begin{equation}
    (\overline{u}_e \gamma_\alpha (1-\gamma^5) v_\nu)(\overline{u}_e \gamma_\beta (1-\gamma^5) v_\nu)^\dagger = 2\mbox{Tr}[\gamma_\alpha \hat{\slashed{k}}_\nu \gamma_\beta (\slashed{a}_e + \slashed{c}_e)(1+\gamma^5)]\, .
\end{equation} 
Taking the traces and dropping terms that vanish after doing the neutrino angular integration, this takes a compact form.
\begin{equation}
    \begin{split}
        \mathcal{M}_{\rm red} &= 64[(g_V+g_A)^2 M^*_{Ln} (a_e+c_e) \cdot (a_p+c_p) +(g_V-g_A)^2 (a_p^0-c_p^0) \tilde{k}_n \cdot (a_e+c_e) \\
        &+M (g_V^2-g_A^2)(\vec{d}_p \cdot (\vec{a}_e+\vec{c}_e)-b_p^0 (a_e^0+c_e^0)]
    \end{split}
\end{equation}
For the final matrix element, use the following shorthand.
\begin{equation}
    [e^\pm] = 1 \pm \frac{k_{ze}}{E_e}, \,
    [p_z^\pm] = \left(1 \pm \frac{k_{zp}}{M^*_{Lp}+M} \right)^2, \,
    [p_B] = \left( \frac{\sqrt{2n_peB}}{M^*_{Lp}+M} \right)^2
\end{equation}
Filling in the final values for $a$--$d$ and dropping terms that go to zero after integrating over $\phi_n$, gives the final matrix element with the argument of all the $I_{n_1, n_2}$ functions being $k_{\perp n}^2/2eB$.  For spin up protons:
\begin{equation}
    \begin{split}
        \mathcal{M}_{\rm red}^\uparrow &= 16 (g_V+g_A)^2 M^*_{Ln} \bigg[ I_{n_e,n_p}^2 [p_z^-] [e^+]  +I_{n_e-1,n_p-1}^2 [p_B] [e^-] \\
        & +2I_{n_e,n_p} I_{n_e-1,n_p-1} \left(1 - \frac{k_{zp}}{M^*_{Lp}+M} \right) \frac{2eB\sqrt{n_e n_p}}{(M^*_{Lp}+M)E_e} \bigg] \\
        &+8(g_V-g_A)^2 \bigg[(M^*_{Ln} - k_{zn} )(I_{n_e,n_p}^2 [p_z^+][e^+] +I_{n_e,n_p-1}^2 [p_B][e^+] \\
        &+ (M^*_{Ln} + k_{zn}) (I_{n_e-1,n_p}^2 [p_z^+][e^-] + I_{n_e-1,n_p-1}^2 [p_B][e^-]) \bigg] \\
        &+ 16 (g_V^2 - g_A^2) M \bigg[ I_{n_e,n_p}^2 \left(-1+\frac{k_{zp}^2}{(M^*_{Lp}+M)^2}\right) [e^+] + I_{n_e-1,n_p-1}^2 [p_B][e^-] \\
        &- 2I_{n_e,n_p}I_{n_e-1,n_p-1} k_{zp} \frac{2eB\sqrt{n_e n_p}}{(M^*_{Lp}+M)^2 E_e} \bigg]
    \end{split}
\end{equation}
For spin down protons:
\begin{equation}
    \begin{split}
        \mathcal{M}_{\rm red}^\downarrow &= 16 (g_V+g_A)^2 M^*_{Ln} \bigg[ I_{n_e,n_p}^2 [p_B] [e^+]  +I_{n_e-1,n_p-1}^2 [p_z^+] [e^-] \\
        & +2I_{n_e,n_p} I_{n_e-1,n_p-1} \left(1 + \frac{k_{zp}}{M^*_{Lp}+M} \right) \frac{2eB\sqrt{n_e n_p}}{(M^*_{Lp}+M)E_e} \bigg] \\
        &+8(g_V-g_A)^2 \bigg[(M^*_{Ln} - k_{zn} )(I_{n_e,n_p}^2 [p_B][e^+] +I_{n_e,n_p-1}^2 [p_z^-][e^+] \\
        &+ (M^*_{Ln} + k_{zn}) (I_{n_e-1,n_p}^2 [p_B][e^-] + I_{n_e-1,n_p-1}^2 [p_z^-][e^-]) \bigg] \\
        &+ 16 (g_V^2 - g_A^2) M \bigg[I_{n_e,n_p}^2 [p_B] [e^+] + I_{n_e-1,n_p-1}^2 \left(-1+\frac{k_{zp}^2}{(M^*_{Lp}+M)^2}\right) [e^-] \\
        &+ 2I_{n_e,n_p}I_{n_e-1,n_p-1} k_{zp} \frac{2eB\sqrt{n_e n_p}}{(M^*_{Lp}+M)^2 E_e} \bigg]
    \end{split}
\end{equation}
Integrating over momenta and adding normalizations gives the total neutrino emissivity, doubled to account for the reverse process.
\begin{equation}
\begin{split}
    Q &= 2 \, \frac{eB G_F^2 \cos^2\theta_c}{4\pi} \sum_{s_p = \{\uparrow, \downarrow\}} \sum_{n_e, n_p} \int \frac{d^3k_\nu}{2E_\nu (2\pi)^3} \frac{d^3k_n}{2M^*_{Ln} (2\pi)^3} \frac{dk_{zp}}{2M^*_{Lp} (2\pi)} \frac{dk_{ze}}{2E_e (2\pi)} \\
    &\times (M^*_{Lp} + M)E_e E_\nu \, n_{FD}(E_n - \mu_n) \, n_{FD}(\mu_p-E_p) \, n_{FD} (\mu_e-E_e) \mathcal{M}_{\rm red}^{(s_p)} \\
    &\times (2\pi)^2 \delta (E_n - E_p - E_e - E_\nu) \delta (k_{zn} - k_{zp} - k_{ze} - k_{z\nu}) 
\end{split}
\end{equation}
where $n_{FD}(E - \mu)= (\mbox{exp}[(E-\mu)/T]+1)^{-1}$ is the Fermi-Dirac distribution. The angular integration is done in a straightforward way by noticing that $k_\nu$ is much smaller than all the other momenta and removing it from the momentum $\delta$-function.  We add an explicit sum over the sign of $k_{zp}$ and $k_{ze}$ and set the limits of integration to $[0 , \infty)$.  $\Theta (x)$ is a step function that is one if $x \geq 0$ and zero otherwise.
\begin{equation}
    \sum_{k_z \rm signs} \int d\Omega_n \, d\Omega_\nu \, \delta (k_{zn} - k_{zp} - k_{ze} ) = \frac{8\pi^2}{k_n} \sum_{k_z \rm signs} \Theta( k_n - |k_{zp} + k_{ze}|)
\end{equation}
The $\delta$-function is used to set the value of $k_{zn}$ which along with $k_{\perp n}$ in the matrix element should be set based on the values of $k_{zp}$ and $k_{ze}$.  Note that in the process of this integration, we did free integrals $d\Omega_\nu$ and $d\phi_n$.  These cause many terms in the matrix element to cancel, which we have already accounted for.

\subsection{Levels of approximation}
\label{sec:approx}

Calculating the phase space integral has the first major deviation from the standard DU calculation.  We must calculate
\begin{equation}
    Q = \frac{eBG_F^2 \cos^2 \theta_c}{256\pi^5} \sum_{n_e, n_p} \sum_{s_p=\pm} \frac{M^*_{Lp}+M}{M^*_{Lp}} \Phi \, ,
\end{equation}
where $\Phi$ is given by
\begin{equation}
\label{eq:ps_integral}
\begin{split}
    \Phi(n_e, n_p, s_p) &= \int \frac{k_n d|k_n|}{M^*_{Ln}} \, E_\nu^3 \, dE_\nu \, d|k_{zp}| \, d|k_{ze}| \, n_{FD}(E_n - \mu_n) \, n_{FD}(\mu_p-E_p)  \\
    & \times n_{FD}(\mu_e -E_e) \delta (E_n - E_p - E_e - E_\nu) \, \mathcal{M}_{\rm red}^{(s_p)} \, .
\end{split}
\end{equation}

For $B\lesssim 10^{16} \, \mbox{G}$, LL are spaced closely together and a quasiclassical (QC) approximation can be made, disregarding effects of LQ~\cite{Baiko_Yakovlev_1999}. The emissivity is given by $Q = R_B Q_\nu^0$ where $Q^\nu_0$ is the emissivity for zero magnetic field given by \cite{yakovlev_kaminker_cooling_formulas}
\begin{equation}
    Q_\nu^0 = \frac{457\pi G_F^2 \cos^2 \theta_c (1+3g_A^2)}{10080} M_{Ln}^* M_{Lp}^* \mu_e T^6
\end{equation}
and $R_B$ quantifies the amount of suppression from being in the ``forbidden region" where DU is not normally allowed. In the forbidden region, the QC suppression $R^{\rm qc}_B$ is given by \cite{Baiko_Yakovlev_1999}
\begin{align}
    R_B^{\rm qc} &= 2 \int d \cos \theta_p \, d \cos \theta_e \frac{k_{Fp} k_{Fe}}{4eB} I_{n_p, n_e}^2 \Theta (k_{Fn} - |k_{Fp} \cos \theta_p + k_{Fe} \cos \theta_e |) \\
    &\approx 2^{-2/3} \int_{-\infty}^\infty ds \int_0^{\pi} d\theta \, \sin^{2/3}\theta \, \text{Ai}^2\left(\frac{x+s^2}{2^{4/3}\sin^{2/3}\theta}\right) \, ,
\end{align}
where $x=[k_{Fn}^2-(k_{Fp}+k_{Fe})^2]/(k_{Fp}^2 N_{Fp}^{-2/3})$ quantifies how far the density is from the DU threshold. Equation~(22) of Ref.~\cite{Baiko_Yakovlev_1999} gives useful approximate expressions for this integral that we will use when comparing to our results. 

To do better, the integral can be performed for each LL separately. Following the spirit of the standard calculation of the DU rate without a magnetic field (see, for example, the appendix of Ref.~\cite{ShapiroTeukolsky1983}), the Fermi surface approximation can be made by setting explicit factors of the energy and momenta as being on the Fermi surface in the second expression.
\begin{equation}
\label{eq:fsa}
    \int d|k_{zp}| \, d|k_{ze}| \, n_{FD} (-E_p) n_{FD}(-E_e)  \rightarrow T^2 \frac{M^*_{Lp} \mu_e}{|k_{Fze} k_{Fzp}|} \int dx_p \, dx_e \, \frac{1}{e^{-x_e} + 1} \frac{1}{e^{-x_p} + 1}
\end{equation}
where $x_i = (E_i-\mu_i)/T$. This presents an obvious problem: this expression is infinite at the exact energy where a new LL becomes available, since at that energy $k_z = 0$ in the highest LL on the Fermi surface.  This occurs because the density of states has a resonance at these energies. Near the resonance, $k_z$ is small and a large range of $k_z$ corresponds to energies within $T$ of the Fermi surface. Doing the full integral gives an enhancement from the resonance in the density of states. In the highest LLs, finite temperature effects are important, even if $T \ll \varepsilon_F$, since $T$ may be of the same order as $k_z^2/E$. Additionally, the functions $I_{n_e,n_p}$ oscillate rapidly as a function of $k_{\perp n}^2/2eB$ for large $n_e$ and $n_p$, and applying a Fermi surface approximation to the argument of $I_{n_e, n_p}$ in order to avoid the computationally costly repeated evaluation of the modified Laguerre polynomial introduces errors. 

These resonances were observed in Refs.~\cite{MARUYAMA2022136813, tambe2024effectmagneticfieldsurca} when calculating the DU rate at low temperature and in Refs.~\cite{Duan_qian_2004, duan_qian_2005_appdx_has_integrals} when calculating the neutrino cross section in low density matter and neglecting the nucleon momentum. Including the nucleon momentum is equivalent in our case to doing the full finite temperature momentum integral, smearing out the resonance due to thermal effects.

An analytical approximation to \(\Phi\) can be used based on the substitution in Eq.~\eqref{eq:fsa} when the ratios \(k^2_{Fze}/\mu_e T\) and \(k^2_{Fzp}/M T\) are both not too small for a given pair of LLs. Continuing with the Fermi surface approximation and performing the integrals, the emissivity for a specific LL is given by
\begin{equation}
\label{eq:sa approx}
    Q^{\rm FSA} (n_e, n_p) = eBG_F^2 \cos^2 \theta_c T^6 \frac{457 \pi}{1290240} \sum_{s_p = \{ \uparrow, \downarrow \}} \frac{(M^*_{Lp} + M) \mu_e}{|k_{Fzp} k_{Fze}|} \sum_{k_z \rm signs} \mathcal{M_{\rm red}} \, .
\end{equation}
So long as the following conditions are met, this approximation is robust.
\begin{equation}
    \cos \theta_p \gg \frac{\sqrt{2M_N T}}{n_p^{1/3}} , \, \,
    \cos \theta_e \gg \sqrt{\frac{T}{n_e^{1/3}}}
\end{equation}
In what follows, we will use this result for LL with the simple prescription that both $\cos \theta_e$ and $\cos \theta_p$ are greater than $0.2$ and neither the electron nor the proton are in their highest two LLs, giving a speed boost of a factor of 5--10. We use these values because the maximum value that the quantity $\sqrt{2M_N T}/n_p^{1/3}$ takes for $T=100 \, \mbox{keV}$ is about $0.16$. The condition that the highest two LL be calculated numerically catches edge cases of poor accuracy at low density. These choices ensure that the dominant contributions to the emissivity (which come from LLs with small \(k_{Fz}\)) are computed in full, while speeding up the computation in parameter space that affects the final result less. If a greater speed boost is desired, a more precise implementation of these conditions can be used. We call this the semi-analytic (SA) approximation. The error introduced by the SA approximation is generally low, particularly for low temperatures, and can be made arbitrarily small by adjusting the tolerance lower. The choice of how stringently to use the SA approximation depends on the desire for accuracy versus computational speed. 

For the highest few LL, the full integral must be calculated. The integral over the neutron energy can be done analytically.
\begin{equation}
    \int_{x_e+x_p}^\infty dx_n \frac{(x_n - x_e - x_p)^3}{e^{x_n}+1} = -6 \ \Li_4(-e^{-x_e-x_p})
\end{equation}
\(\Li_4(z)\) is the polylogarithm of order 4, defined by \(\sum_{k=1}^\infty z^k/k^4\) where it converges and its analytic continuation elsewhere. Making the substitution \(k_{zi} \rightarrow \bar{k}_{zi} \equiv k_{zi} / T\) gives
\begin{equation}
   \Phi = -6 T^6 \int d\bar{k}_{ze} d\bar{k}_{zp} n_{FD}(-x_e)n_{FD}(-x_p) \Li_4(-e^{-x_e-x_p}) \Theta(k_{Fn} - |k_{zp} + k_{zn}|) \mathcal{M}_{\rm red}^{(s)} \, .
\end{equation}

\begin{figure}[tb]
    \includegraphics[width=0.8\textwidth]{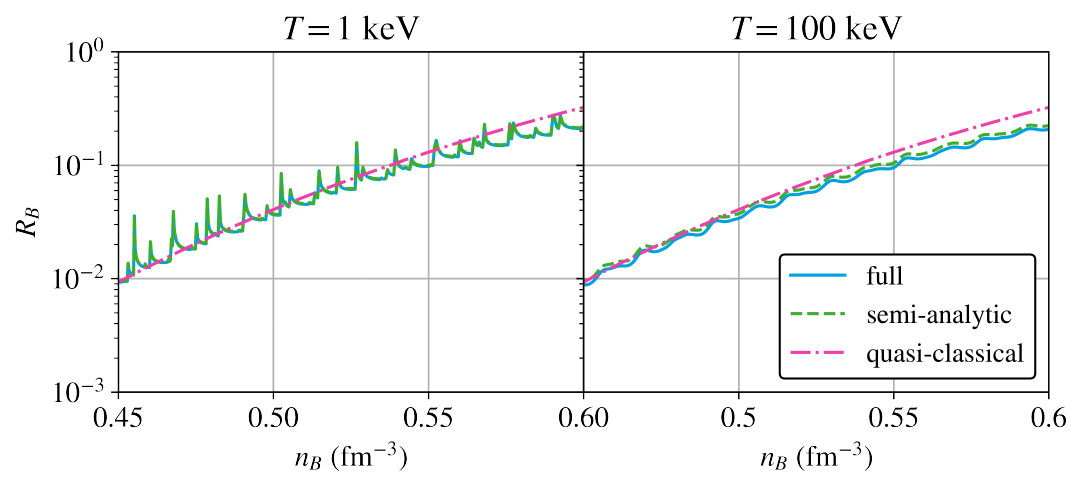}
    \caption{$R_B$ comparing the semi-analytic and quasi-classical approximations with the full calculation just below the DU threshold at $B = 5 \times 10^{16}$ G. In the left panel the semi-analytic approximation and full calculation are indistinguishable.}
    \label{fig:rb}
\end{figure}

This integral can be found numerically. Figure~\ref{fig:rb} compares $R_B$ in the full calculation with the SA and QC approximations for $B=5 \times 10^{16} \, \mbox{G}$ at $T = 1$ keV and $100$ keV.  Clearly visible at low temperature are the densities at which new LLs become available and there is an enhancement to the emissivity. Since the proton has a large anomalous magnetic moment, there are separate peaks when new LLs become available for electrons and for spin up and spin down protons. The deviation from the QC calculation at high density is due to relativistic corrections that we include that the QC approximation neglects.

Different LL for the proton are spaced much more closely together in energy than for the electron because of the large mass of the proton. When the temperature is of the same order as the energy splitting $T \simeq eB/M$, peaks in the emissivity due to new proton LL become are thermally smeared. For a magnetic field of $10^{16} \, \mbox{G}$ this becomes an important effect at $T \gtrsim 100 \, \mbox{keV}$ and LL above the Fermi surface should be included. In contrast, the energy splitting of the electron LL is $\sqrt{eB/4n_e}$ for $n_e \gg 1$. Thermal smearing of electrons becomes important for $n_e \leq 100$ and $B = 10^{16} \, \mbox{G}$ only when $T \gtrsim 1 \, \mbox{MeV}$. Once the star is more than a few seconds old, thermal smearing of electron LL can be neglected for such strong fields. 

\subsection{Solutions to computational challenges}
\label{sec:computing}

Computing \(I_{n,r}(x)\) is the main bottleneck for the calculation of the full emissivity (and later the opacity). A significant improvement can be gained by using the identity
\begin{equation}
    L_r^{n-r}(x) \rightarrow {n \choose r} M(-r, n-r+1, x)
\end{equation}
to remove the need for recursive computation of Laguerre polynomials. To get a more significant improvement for modest densities, we pre-compute $I_{n_e, n_p} (x)$ for $n_e, n_p \leq 250$. For a given pair of indices \(n_e\) and \(n_p\), \(I_{n_e,n_p}(k_{\perp n}^2/2eB)\) is mostly confined to the region
\begin{equation}
    2eB (\sqrt{n_e} - \sqrt{n_p})^2 < k_{\perp n}^2 < 2eB (\sqrt{n_e} + \sqrt{n_p})^2 \, .
\end{equation}
Far above and below this range, the function $I_{n_e, n_p} (k_{\perp n}^2/2eB)$ is suppressed. This condition is equivalent to the requirement that the electron and proton wavefunctions contain transverse momentum components of the same order as the transverse momentum of the neutron. When in the forbidden region, the integral solely samples values of $k_{\perp n}$ near the boundaries of this region and values should be precomputed somewhat beyond these limits. Additionally, $I_{n_e, n_p}$ has $n_e + 1$ extrema if $n_e \leq n_p$ and $n_p + 1$ extrema if $n_p < n_e$. If one of $n_e$ or $n_p$ is small, relatively few points need to be pre-computed to capture the full functional dependence on $k_{\perp n}$.

Based on these observations, the final computation of \(I_{n_e,n_p}\) uses a lookup table, with values of $k_{\perp n}$ in the relevant region and precision scaling with the number of extrema. Between the precomputed points, quadratic Lagrange interpolation approximates the value of the function. This reduces the runtime of the function from tens of microseconds to 200-300 nanoseconds, with relative error less than 0.1\% for almost all values of $k_{\perp n}$.

\subsection{Emissivity results}
\label{sec:emissivity results}

To understand the relevance of our results to the thermal evolution of NSs, we compare the Direct Urca emissivity calculated in the QC approximation, in our SA approximation, and by calculating the full integral. Where relevant, we compare these emissivities to the cooling rates calculated for Modified Urca and bremsstrahlung processes calculated with zero magnetic field.

\begin{figure}[tb]
    \includegraphics[width=\textwidth]{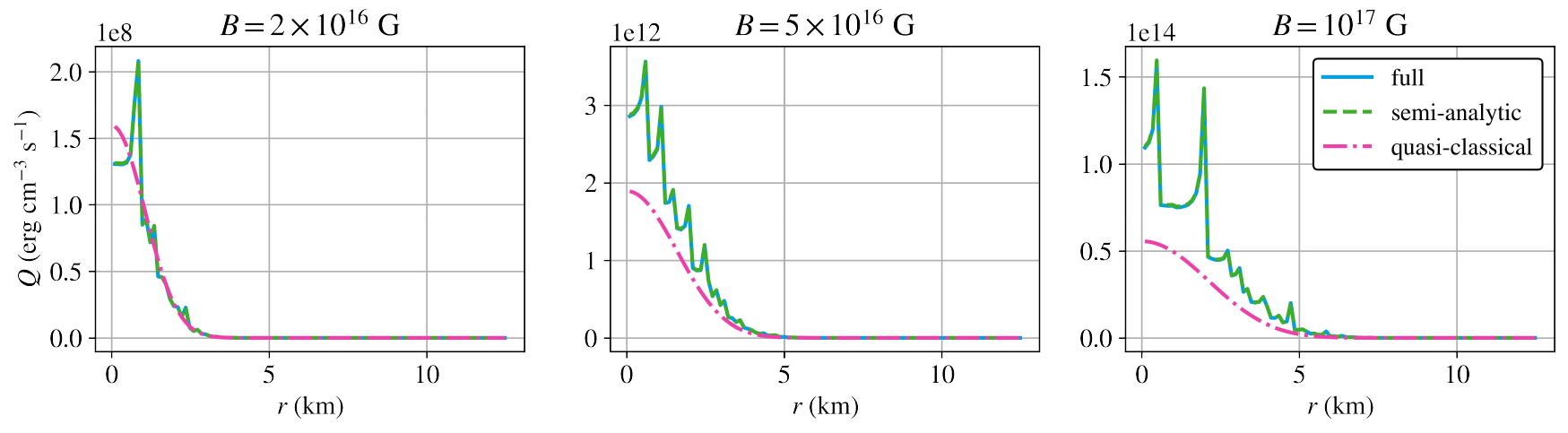}
    \caption{Radial profiles of the DU emissivity for various magnetic field strengths in a NS with \(M = 1.4M_\odot\) and redshifted temperature \(\tilde{T} = 1 \text{ keV}\). At this temperature, the semi-analytic approximation and full calculation are indistinguishable.}
    \label{fig:star_Q_radius_1kev}
\end{figure}

\begin{figure}[tb]
    \includegraphics[width=\textwidth]{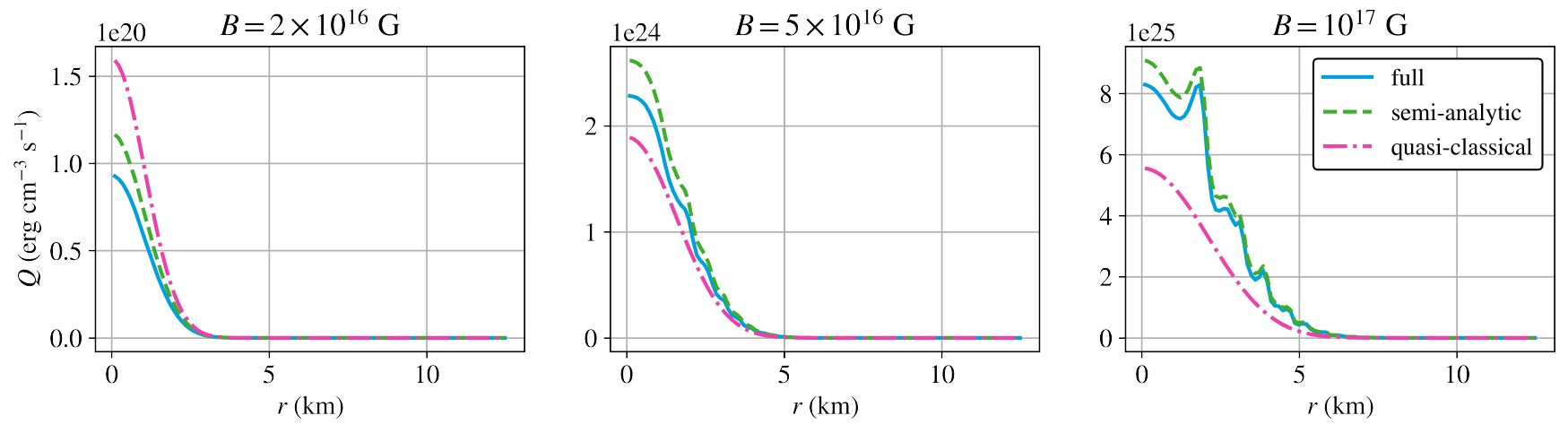}
    \caption{Radial profiles of the DU emissivity for various magnetic field strengths in a NS with \(M = 1.4M_\odot\) and redshifted temperature \(\tilde{T} = 100 \text{ keV}\).}
    \label{fig:star_Q_radius_100kev}
\end{figure}

Figures~\ref{fig:star_Q_radius_1kev} and \ref{fig:star_Q_radius_100kev} show the emissivity as a function of radius for a $1.4 \, M_\odot$ NS with a redshifted temperature $\tilde{T} = 1 \, \mbox{keV}$ and $100 \, \mbox{keV}$ respectively with various choices of magnetic field. We consider a $1.4M_\odot$ NS as a representative example where effects below the Direct Urca threshold are important as, in our equation of state, the entire star is below the Direct Urca threshold. Integrated total emissivity is given in Table~\ref{tab:emissivity}. For a field strength of $2 \times 10^{16} \, \mbox{G}$, slow cooling is many orders of magnitude more efficient than Direct Urca while for larger fields DU dominates. At $B = 2 \times 10^{16} \, \mbox{G}$, our results are parametrically lower than the QC approximation because we include relativistic corrections while the QC approximation is non-relativistic. At larger magnetic fields, we find that the QC approximation somewhat underestimates the emissivity but is unexpectedly robust even when only tens of LL are occupied. Since relativistic corrections in general decrease the emissivity we find, a non-relativistic calculation of the full emissivity would find a somewhat larger deviation from the QC approximation, but the results would still be within the same order of magnitude.

Table~\ref{tab:emissivity} gives the total emissivity of a $1.4 \, M_\odot$ NS for a variety of temperatures and magnetic fields. Since the peaks due to resonances are averaged over, the enhancement from the full calculation is modest. For the purposes of a simulation of an entire NS to predict its observed surface temperature, this is probably not an important correction. Only if some phenomenon hinges on the thermal behavior of the star at specific densities does this effect become important. It is possible that transport coefficients are one such parameter, with Urca processes being a source of viscosity in neutron star matter \cite{Harris_urcaviscosity}. The possibility of a non-monotonic viscosity sourced by resonant Urca processes may have implications for understanding the oscillations of magnetars and should be further studied.

\begin{table}[tb]
\renewcommand{\arraystretch}{1.2}
\setlength{\tabcolsep}{15pt}
\caption{Total emissivity (erg/s) of a $1.4 \, M_\odot$ NS for Direct Urca calculated in full, in the SA approximation, and in the QC approximation in the presence of a strong magnetic field. Slow cooling (Modified Urca and bremsstrahlung) with zero magnetic field shown for comparison.}
\begin{tabular}{|c|c||c|c|c|c|}
\hline
B ($10^{16} \, \mbox{G}$) & $\tilde{T}$ (keV) & Full DU & SA DU & QC DU & Slow ($B=0$)\\
\hline
$2$ & $1$ & \(2.83 \times 10^{24}\) &  \(2.84 \times 10^{24}\) & \(2.74 \times 10^{24}\) & \(5.40 \times 10^{24}\) \\ \hline
$2$ & $100$ & \(1.65 \times 10^{36}\) & \(2.05 \times 10^{36} \)& \(2.74 \times 10^{36}\) & \(5.40 \times 10^{40}\) \\ \hline
$5$ & $1$ & \(1.67 \times 10^{29}\) & \(1.68 \times 10^{29}\) & \(1.05 \times 10^{29}\) & \(5.40 \times 10^{24}\) \\ \hline
$5$ & $100$ & \(1.35 \times 10^{41}\) & \(1.56 \times 10^{41}\) & \(1.05 \times 10^{41}\) & \(5.40 \times 10^{40}\) \\ \hline
$10$ & $1$ & \(1.48 \times 10^{31}\) & \(1.49 \times 10^{31}\) & \(6.89 \times 10^{30}\) & \(5.40 \times 10^{24}\) \\ \hline
$10$ & $100$ & \(1.33 \times 10^{43}\) & \(1.45 \times 10^{43}\) & \(6.89 \times 10^{42}\) & \(5.40 \times 10^{40}\) \\ \hline
\end{tabular}
\label{tab:emissivity}
\end{table}

\section{Neutrino Opacities}
\label{sec:opacity}

\subsection{Low density conditions}

The same tools we use to calculate the DU emissivity can be applied to calculate the opacity for neutrinos to capture on neutrons ($\nu + n \rightarrow e + p$) and for antineutrinos to capture on protons ($\bar{\nu} + p \rightarrow e^+ + n$). In the ejecta from a NS merger, the neutrino decoupling region ranges from $T\simeq 10 \, \mbox{MeV}$ and $n \simeq 0.1 \, n_{\rm sat}$ for soft neutrinos ($E_\nu \sim 3 \, \mbox{MeV}$) to $T \simeq 2 \, \mbox{MeV}$ and $n \simeq 10^{-5} \, n_{\rm sat}$ for harder neutrinos ($E_\nu \sim 50 \, \mbox{MeV}$)~\cite{Endrizzi_2020}. Magnetohydrodynamic instabilities in the merger remnant can amplify the magnetic field of the merging NSs by many orders of magnitude, possibly larger than $10^{16} \, \mbox{G}$~\cite{PRD_mergerfields, Science_PriceRosswog}. We calculate modifications to the neutrino opacity in the merger environment due to magnetic fields. This is similar to previous calculations performed in Refs.~\cite{Duan_qian_2004, duan_qian_2005_appdx_has_integrals} where minimal deviation from the zero field result was found, but we will consider stronger magnetic fields and higher densities where Maxwell-Boltzmann statistics they utilize are no longer appropriate and final state blocking becomes relevant.

The opacity for neutrinos capturing on neutrons not including the effects of stimulated absorption is given by
\begin{equation}
\label{eq:kappan}
    \begin{split}
        \kappa_{\nu n} &= \frac{G_F^2 \cos^2 \theta_c eB}{64\pi^3} \sum_{\substack{s_n, s_p\\ n_e, n_p}} \int dk_{ze} \, dk_{zp} \, E_n \Theta (k_n - |k_{zp} + k_{ze} - k_{z\nu}|) \mathcal{M}_{\rm red}^{(s_n, s_p)} \\
        &\times n_{FD}(E_n - \mu_n) n_{FD} (\mu_p - E_p) n_{FD} (\mu_e - E_e) \, . 
    \end{split}
\end{equation}
To include stimulated absorption, this quantity should be multiplied by $1/(1-\mathcal{F}'_\nu)$ for $\mathcal{F}_\nu$ the invariant distribution function for neutrinos that need not be in chemical equilibrium. Note that we are using a different normalization for the reduced matrix element at low density so that it is dimensionless and does not have an internal sum over the neutron spin since at low density the neutron anomalous magnetic moment is important. The opacity for antineutrinos capturing on protons is given by
\begin{equation}
\label{eq:kappap}
    \begin{split}        
        \kappa_{\bar{\nu} p} &= \frac{G_F^2 \cos^2 \theta_c eB}{64\pi^3} \sum_{\substack{s_n, s_p\\ n_e, n_p}} \int dk_{ze} \, dk_{zp} \, E_n \Theta (k_n - |k_{zp} + k_{z\nu} - k_{ze}|) \mathcal{M}_{\rm red}^{(s_n, s_p)} \\
        &\times n_{FD}(\mu_n - E_n) n_{FD}(E_p - \mu_p) n_{FD}(-\mu_e - E_e) \, .
    \end{split}
\end{equation}
For simplicity, we expand the matrix element to zeroth order in the momentum of the nucleons since we are considering matter well below saturation density. As noted in Ref.~\cite{duan_qian_2005_appdx_has_integrals}, one cannot neglect the nucleon momentum when calculating chemical equilibrium and energy conservation without producing spurious infinities, even at low density. However, the matrix element can safely be expanded to low order in the momentum at low density since the dependence is linear in $k/M$. At this order, and within the approximation $E_\nu \ll \sqrt{2 M T} \simeq k_n$, the matrix elements for both processes are the same and are given by the following \cite{duan_qian_2005_appdx_has_integrals}.
\begin{equation}
\begin{split}
    \mathcal{M}_{\rm red}^{(s_n = +, s_p = + )} & = 2(g_V + g_A)^2 \left( 1 + \frac{k_{ze}}{E_e} \right) (1 + \cos \theta_\nu) I_{n_e, n_p}^2 \\
    &+ 2(g_V - g_A)^2 \left( 1 - \frac{k_{ze}}{E_e} \right) (1 - \cos \theta_\nu) I_{n_e - 1, n_p}^2 \\
    \mathcal{M}_{\rm red}^{(s_n = +, s_p = -)} & = 8g_A^2 \left( 1 - \frac{k_{ze}}{E_e} \right) (1 + \cos \theta_\nu) I_{n_e-1, n_p-1}^2\\
    \mathcal{M}_{\rm red}^{(s_n = -, s_p = +)} & = 8g_A^2 \left( 1 + \frac{k_{ze}}{E_e} \right) (1 - \cos \theta_\nu) I_{n_e, n_p}^2\\
    \mathcal{M}_{\rm red}^{(s_n = -, s_p = -)} & = 2(g_V + g_A)^2 \left( 1 - \frac{k_{ze}}{E_e} \right) (1 - \cos \theta_\nu) I_{n_e-1, n_p-1}^2 \\
    &+ 2(g_V - g_A)^2 \left( 1 + \frac{k_{ze}}{E_e} \right) (1 + \cos \theta_\nu) I_{n_e, n_p-1}^2
\end{split}
\end{equation}

\begin{figure}[tb]
    \includegraphics[width=0.9\textwidth]{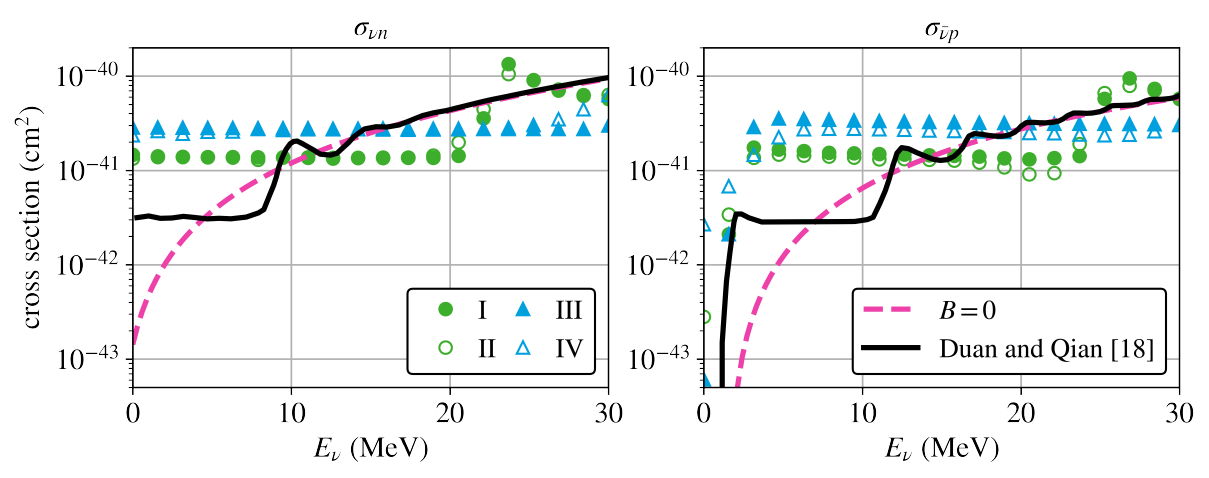}
    \caption{Cross sections for neutrinos to capture on nucleons. Green circles correspond to $B = 5 \times 10^{16} \, \mbox{G}$ while blue triangles are for $B = 10^{17} \, \mbox{G}$. Filled marks are at $T = 1 \, \mbox{MeV}$ while unfilled are at $T = 3\, \mbox{MeV}$ and $8 \, \mbox{MeV}$, respectively. (See Table~\ref{tab:conditions}) The black curve for comparison was calculated in Ref.~\cite{duan_qian_2005_appdx_has_integrals} at $B = 10^{16}$ G and $T = 2$ MeV.}
    \label{fig:cross_section}
\end{figure}

To compare with the literature, we also calculate the cross sections for these processes using Maxwell-Boltzmann statistics for the target nucleon. Figure~\ref{fig:cross_section} shows the cross section for these two processes (given by the same integration as Eqs.~\eqref{eq:kappan} and \eqref{eq:kappap}, without final state blocking and dividing by the target nucleon density) for neutrinos propagating perpendicular to the magnetic field. The orientation of the neutrino momentum was found in Refs.~\cite{Duan_qian_2004, duan_qian_2005_appdx_has_integrals} to have a very small effect on the neutron branch and almost no effect on the proton branch. As a benchmark, we compare with the cross sections calculated in Ref.~\cite{duan_qian_2005_appdx_has_integrals} at $B = 10^{16} \, \mbox{G}$ and $T = 2 \, \mbox{MeV}$ and the cross section with no magnetic field. Table~\ref{tab:conditions} lists the choices of magnetic field and temperatures we use. At low neutrino energy, the cross section is strongly enhanced by the magnetic field because of the contribution of the anomalous magnetic moment of the neutron shifting the effective mass splitting of the nucleons. For sufficiently large fields, the cross section for capture on protons does not go to zero even for zero neutrino energy because the contribution of the anomalous magnetic moment is of the same order as $m_n + m_e - m_p \simeq 1.8 \, \mbox{MeV}$.

As expected, the first resonance occurs at higher neutrino energy and enhances the cross section by a much larger factor than at smaller magnetic field. This warrants further inquiry, particularly at a larger range of densities, temperatures, and with the necessary additional angular integrations to safely consider higher neutrino energies. The results of Refs.~\cite{Duan_qian_2004, duan_qian_2005_appdx_has_integrals} indicate that at large neutrino energy, the opacities should approach the value with zero field with small corrections. Whether the highest energy neutrinos from a neutron star merger are in this regime for such strong fields is unclear.

\begin{table}[tb]
\renewcommand{\arraystretch}{1.2}
\setlength{\tabcolsep}{15pt}
\caption{Conditions in which we calculate the opacity}
\begin{tabular}{|c||c|c|}
\hline
& $B$ (G) & $T$ (MeV) \\
\hline
I & $5 \times 10^{16}$ & 1 \\
\hline
II & $5 \times 10^{16}$ & 3 \\
\hline
III & $10^{17}$ & 1 \\
\hline
IV & $10^{17}$ & 8 \\
\hline
\end{tabular}
\label{tab:conditions}
\end{table}

\subsection{Opacity results}
\label{sec:opacity results}

\begin{figure}[tb]
    \includegraphics[width=0.9\textwidth]{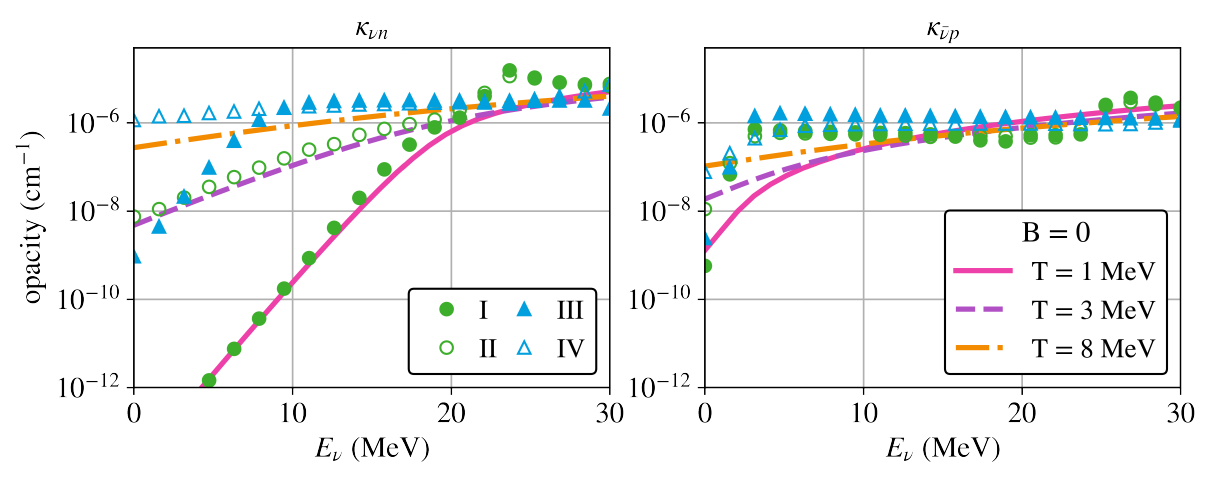}
    \caption{Neutrino opacities at $n_B = 0.001 \, n_{\rm sat}$ and $Y_p = 0.25$. Green circles correspond to $B = 5 \times 10^{16} \, \mbox{G}$ while blue triangles are for $B = 10^{17} \, \mbox{G}$. Filled marks are at $T = 1 \, \mbox{MeV}$ while unfilled are at $T = 3\, \mbox{MeV}$ and $8 \, \mbox{MeV}$, respectively. (See Table~\ref{tab:conditions}) Curves show results calculated at zero magnetic field.}
    \label{fig:kappa_low}
\end{figure}

\begin{figure}[tb]
    \includegraphics[width=0.9\textwidth]{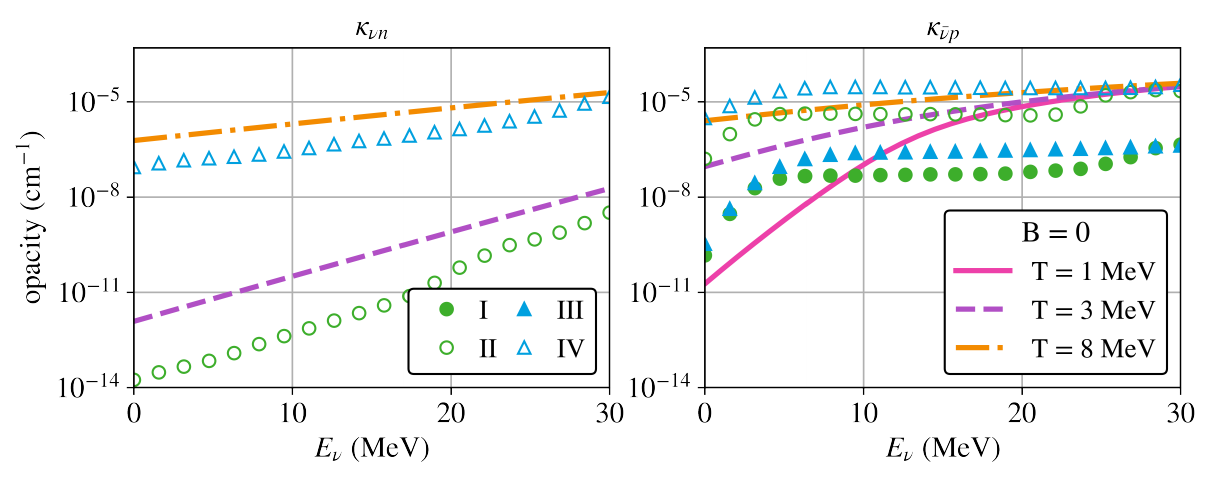}
    \caption{Neutrino opacities at $n_B = 0.1 \, n_{\rm sat}$ and $Y_p = 0.1$. Green circles correspond to $B = 5 \times 10^{16} \, \mbox{G}$ while blue triangles are for $B = 10^{17} \, \mbox{G}$. Filled marks are at $T = 1 \, \mbox{MeV}$ while unfilled are at $T = 3\, \mbox{MeV}$ and $8 \, \mbox{MeV}$, respectively. (See Table~\ref{tab:conditions}) Curves show results calculated at zero magnetic field. At $T = 1 \, \mbox{MeV}$, the opacity for capture on neutrons is strongly suppressed with or without a magnetic field due to being far out of beta equilibrium.}
    \label{fig:kappa_high}
\end{figure}

We calculate $\kappa_{\nu n}$ and $\kappa_{\bar{\nu} p}$ at two densities in the ejecta, $0.1 \, n_{\rm sat}$ and $0.001 \, n_{\rm sat}$ with proton fraction $Y_p = 0.1$ and $Y_p = 0.25$ respectively. Figures~\ref{fig:kappa_low} and~\ref{fig:kappa_high} show our results for $\kappa_{\nu n}$ and $\kappa_{\bar{\nu} p}$ at our low and high density respectively and show the results for the opacity at the same densities and temperatures with no magnetic field.

The most important effects the magnetic field has on the opacity for capturing on neutrons is due to suppression of Pauli blocking of electrons at very low density. At $0.001 \, n_{\rm sat}$ and $Y_p = 0.25$ (Fig.~\ref{fig:kappa_low}), a strong magnetic field suppresses the electron chemical potential by a factor of a few because the density is linearly dependent on $\mu_e$ when the number of LL is small, bringing the system much closer to beta equilibrium and suppressing Pauli blocking of electrons. This enhances the opacity by many orders of magnitude (compare Conditions I and III in the left panel of Fig.~\ref{fig:kappa_low}, both at $T = 1 \, \mbox{MeV}$ but with different magnetic field strength). At $0.1 \, n_{\rm sat}$ and $Y_p = 0.1$ (Fig.~\ref{fig:kappa_high}), the electrons are strongly degenerate and the system is far from beta equilibrium, suppressing capture on neutrons even in the presence of a strong magnetic field. The opacity with a magnetic field is suppressed relative to the zero field calculation since the energy available for an electron (approximately $\mu_n - \mu_p + E_\nu$) is only sufficient to populate the lowest few LL. As the neutrino energy is increased, the opacity approaches the zero field value.

For capture on protons, the most important effect for low energy neutrinos is from the magnetic moments of the nucleons. Since this process produces positrons, the leptons are not Pauli blocked. Normally, the opacity for neutrinos with energy $E_\nu < m_e + m_n - m_p$ to capture on protons is suppressed. For the strongest magnetic fields we consider $B \simeq 10^{17} \, \mbox{G}$, the energy contribution of the anomalous magnetic moments of the nucleons are of the same order as the positron mass and the mass splitting of the nucleons and this suppression is lifted. In the presence of a strong magnetic field, neutrinos with energies $E_\nu \lesssim 10 \, \mbox{MeV}$ have their opacity to capture on protons enhanced by orders of magnitude due to this effect. For higher energy neutrinos, the opacity is suppressed just like the capture rate on neutrons is suppressed at $0.1 \, n_{\rm sat}$ due to only being able to access the lowest few LLs. At higher neutrino energies once many LL can be populated, the opacity will approach the zero field value.

\section{Conclusion}
\label{sec:conclusion}

In this work, we calculate the Direct Urca emissivity in the presence of magnetic fields $B \geq 2 \times 10^{16} \, \mbox{G}$ including the effects of Landau quantization and full relativity for the V-A weak Lagrangian. We find that relativistic corrections tend to suppress the emissivity, but that the often utilized quasiclassical approximation of Ref.~\cite{Baiko_Yakovlev_1999} underestimates the emissivity for fields $B \geq 5 \times 10^{16} \, \mbox{G}$. We present a semi-analytic approximation, in which analytic results within the Fermi surface approximation given by Eq.~\eqref{eq:sa approx} are used for Landau levels far from resonance and the full phase space integral given by Eq.~\eqref{eq:ps_integral} is calculated for the highest few Landau levels. The semi-analytic approximation captures resonances at specific densities that the quasiclassical approximation misses, especially at low temperatures. When calculating the emissivity of an entire neutron star, we find that the quasiclassical approximation is correct to within less than an order of magnitude even for these very large fields and the high thermal conductivity of the core likely washes out these resonances when considering global thermal evolution. Whether these resonances might have implications for transport warrants further study.

Applying the same techniques, we calculate neutrino opacities in conditions relevant for binary neutron star merger ejecta at magnetic fields $B \geq 5 \times 10^{16} \, \mbox{G}$. We find significant enhancement to the rates of absorption at these strong fields for low energies neutrinos relative to the zero field result. This is due to the large anomalous magnetic moments of the nucleons and suppression of the electron chemical potential by the magnetic field at low density. If small regions of strong magnetic fields develop in the ejecta of a neutron star merger, neutrino capture would be locally enhanced, the neutrinosphere could become distorted, and the proton fraction in these regions would be changed. These results motivate a more exhaustive study of neutrino opacities at superstrong magnetic fields which may have important implications for neutrino transport and nucleosynthesis in binary neutron star merger simulations.

\section*{Acknowledgments}
The U.S. DOE supported the work of M. K. and C. W. under Grant No. DE-FG02-00ER41132. C. W. was additionally supported by the N3AS Physics Frontiers Center's NSF award No. 2020275 and the Washington NASA Space Grant's UW Summer Undergraduate Research Program. We thank Yong-Zhong Qian, David Radice, Peter Rau, Sanjay Reddy, Richard Mikaël Slevinsky, and John Stroud for helpful discussions and suggestions and the INT Undergraduate Research Network for administrative support and useful conversations.

\section*{Data Availability Statement}
The data that support the findings of this article are openly available~\cite{nslq}.

\bibliographystyle{apsrev4-1}
\bibliography{mag_durca_refs}

\begin{thebibliography}{38}%
\makeatletter
\providecommand \@ifxundefined [1]{%
 \@ifx{#1\undefined}
}%
\providecommand \@ifnum [1]{%
 \ifnum #1\expandafter \@firstoftwo
 \else \expandafter \@secondoftwo
 \fi
}%
\providecommand \@ifx [1]{%
 \ifx #1\expandafter \@firstoftwo
 \else \expandafter \@secondoftwo
 \fi
}%
\providecommand \natexlab [1]{#1}%
\providecommand \enquote  [1]{``#1''}%
\providecommand \bibnamefont  [1]{#1}%
\providecommand \bibfnamefont [1]{#1}%
\providecommand \citenamefont [1]{#1}%
\providecommand \href@noop [0]{\@secondoftwo}%
\providecommand \href [0]{\begingroup \@sanitize@url \@href}%
\providecommand \@href[1]{\@@startlink{#1}\@@href}%
\providecommand \@@href[1]{\endgroup#1\@@endlink}%
\providecommand \@sanitize@url [0]{\catcode `\\12\catcode `\$12\catcode
  `\&12\catcode `\#12\catcode `\^12\catcode `\_12\catcode `\%12\relax}%
\providecommand \@@startlink[1]{}%
\providecommand \@@endlink[0]{}%
\providecommand \url  [0]{\begingroup\@sanitize@url \@url }%
\providecommand \@url [1]{\endgroup\@href {#1}{\urlprefix }}%
\providecommand \urlprefix  [0]{URL }%
\providecommand \Eprint [0]{\href }%
\providecommand \doibase [0]{http://dx.doi.org/}%
\providecommand \selectlanguage [0]{\@gobble}%
\providecommand \bibinfo  [0]{\@secondoftwo}%
\providecommand \bibfield  [0]{\@secondoftwo}%
\providecommand \translation [1]{[#1]}%
\providecommand \BibitemOpen [0]{}%
\providecommand \bibitemStop [0]{}%
\providecommand \bibitemNoStop [0]{.\EOS\space}%
\providecommand \EOS [0]{\spacefactor3000\relax}%
\providecommand \BibitemShut  [1]{\csname bibitem#1\endcsname}%
\let\auto@bib@innerbib\@empty
\bibitem [{\citenamefont {Olausen}\ and\ \citenamefont
  {Kaspi}(2014)}]{Olausen_2014_magnetar_catalogue}%
  \BibitemOpen
  \bibfield  {author} {\bibinfo {author} {\bibfnamefont {S.~A.}\ \bibnamefont
  {Olausen}}\ and\ \bibinfo {author} {\bibfnamefont {V.~M.}\ \bibnamefont
  {Kaspi}},\ }\href {\doibase 10.1088/0067-0049/212/1/6} {\bibfield  {journal}
  {\bibinfo  {journal} {The Astrophysical Journal Supplement Series}\ }\textbf
  {\bibinfo {volume} {212}},\ \bibinfo {pages} {6} (\bibinfo {year}
  {2014})}\BibitemShut {NoStop}%
\bibitem [{\citenamefont {Suvorov}\ and\ \citenamefont
  {Melatos}(2023)}]{Suvorov_2023_GLEAM_magnetar}%
  \BibitemOpen
  \bibfield  {author} {\bibinfo {author} {\bibfnamefont {A.~G.}\ \bibnamefont
  {Suvorov}}\ and\ \bibinfo {author} {\bibfnamefont {A.}~\bibnamefont
  {Melatos}},\ }\href {\doibase 10.1093/mnras/stad274} {\bibfield  {journal}
  {\bibinfo  {journal} {Monthly Notices of the Royal Astronomical Society}\
  }\textbf {\bibinfo {volume} {520}},\ \bibinfo {pages} {1590–1600} (\bibinfo
  {year} {2023})}\BibitemShut {NoStop}%
\bibitem [{\citenamefont {Leinson}\ and\ \citenamefont
  {Pérez}(1998)}]{Leinson_perez}%
  \BibitemOpen
  \bibfield  {author} {\bibinfo {author} {\bibfnamefont {L.~B.}\ \bibnamefont
  {Leinson}}\ and\ \bibinfo {author} {\bibfnamefont {A.}~\bibnamefont
  {Pérez}},\ }\href {\doibase 10.1088/1126-6708/1998/09/020} {\bibfield
  {journal} {\bibinfo  {journal} {Journal of High Energy Physics}\ }\textbf
  {\bibinfo {volume} {1998}},\ \bibinfo {pages} {020} (\bibinfo {year}
  {1998})}\BibitemShut {NoStop}%
\bibitem [{\citenamefont {{Baiko}}\ and\ \citenamefont
  {{Yakovlev}}(1999)}]{Baiko_Yakovlev_1999}%
  \BibitemOpen
  \bibfield  {author} {\bibinfo {author} {\bibfnamefont {D.~A.}\ \bibnamefont
  {{Baiko}}}\ and\ \bibinfo {author} {\bibfnamefont {D.~G.}\ \bibnamefont
  {{Yakovlev}}},\ }\href {\doibase 10.48550/arXiv.astro-ph/9812071} {\bibfield
  {journal} {\bibinfo  {journal} {Astronomy \& Astrophysics}\ }\textbf
  {\bibinfo {volume} {342}},\ \bibinfo {pages} {192} (\bibinfo {year}
  {1999})},\ \Eprint {http://arxiv.org/abs/astro-ph/9812071}
  {arXiv:astro-ph/9812071 [astro-ph]} \BibitemShut {NoStop}%
\bibitem [{\citenamefont {Yakovlev}\ \emph {et~al.}(2001)\citenamefont
  {Yakovlev}, \citenamefont {Kaminker}, \citenamefont {Gnedin},\ and\
  \citenamefont {Haensel}}]{yakovlev_kaminker_cooling_formulas}%
  \BibitemOpen
  \bibfield  {author} {\bibinfo {author} {\bibfnamefont {D.}~\bibnamefont
  {Yakovlev}}, \bibinfo {author} {\bibfnamefont {A.}~\bibnamefont {Kaminker}},
  \bibinfo {author} {\bibfnamefont {O.}~\bibnamefont {Gnedin}}, \ and\ \bibinfo
  {author} {\bibfnamefont {P.}~\bibnamefont {Haensel}},\ }\href {\doibase
  https://doi.org/10.1016/S0370-1573(00)00131-9} {\bibfield  {journal}
  {\bibinfo  {journal} {Physics Reports}\ }\textbf {\bibinfo {volume} {354}},\
  \bibinfo {pages} {1} (\bibinfo {year} {2001})}\BibitemShut {NoStop}%
\bibitem [{\citenamefont {Page}\ \emph {et~al.}(2004)\citenamefont {Page},
  \citenamefont {Lattimer}, \citenamefont {Prakash},\ and\ \citenamefont
  {Steiner}}]{Page_2004}%
  \BibitemOpen
  \bibfield  {author} {\bibinfo {author} {\bibfnamefont {D.}~\bibnamefont
  {Page}}, \bibinfo {author} {\bibfnamefont {J.~M.}\ \bibnamefont {Lattimer}},
  \bibinfo {author} {\bibfnamefont {M.}~\bibnamefont {Prakash}}, \ and\
  \bibinfo {author} {\bibfnamefont {A.~W.}\ \bibnamefont {Steiner}},\ }\href
  {\doibase 10.1086/424844} {\bibfield  {journal} {\bibinfo  {journal} {The
  Astrophysical Journal Supplement Series}\ }\textbf {\bibinfo {volume}
  {155}},\ \bibinfo {pages} {623} (\bibinfo {year} {2004})}\BibitemShut
  {NoStop}%
\bibitem [{\citenamefont {Page}\ \emph {et~al.}(2006)\citenamefont {Page},
  \citenamefont {Geppert},\ and\ \citenamefont {Weber}}]{PAGE2006497}%
  \BibitemOpen
  \bibfield  {author} {\bibinfo {author} {\bibfnamefont {D.}~\bibnamefont
  {Page}}, \bibinfo {author} {\bibfnamefont {U.}~\bibnamefont {Geppert}}, \
  and\ \bibinfo {author} {\bibfnamefont {F.}~\bibnamefont {Weber}},\ }\href
  {\doibase https://doi.org/10.1016/j.nuclphysa.2005.09.019} {\bibfield
  {journal} {\bibinfo  {journal} {Nuclear Physics A}\ }\textbf {\bibinfo
  {volume} {777}},\ \bibinfo {pages} {497} (\bibinfo {year} {2006})},\ \bibinfo
  {note} {special Issue on Nuclear Astrophysics}\BibitemShut {NoStop}%
\bibitem [{\citenamefont {Enoto}\ \emph {et~al.}(2019)\citenamefont {Enoto},
  \citenamefont {Kisaka},\ and\ \citenamefont {Shibata}}]{Enoto_2019}%
  \BibitemOpen
  \bibfield  {author} {\bibinfo {author} {\bibfnamefont {T.}~\bibnamefont
  {Enoto}}, \bibinfo {author} {\bibfnamefont {S.}~\bibnamefont {Kisaka}}, \
  and\ \bibinfo {author} {\bibfnamefont {S.}~\bibnamefont {Shibata}},\ }\href
  {\doibase 10.1088/1361-6633/ab3def} {\bibfield  {journal} {\bibinfo
  {journal} {Reports on Progress in Physics}\ }\textbf {\bibinfo {volume}
  {82}},\ \bibinfo {pages} {106901} (\bibinfo {year} {2019})}\BibitemShut
  {NoStop}%
\bibitem [{\citenamefont {Manchester}\ \emph {et~al.}(2005)\citenamefont
  {Manchester}, \citenamefont {Hobbs}, \citenamefont {Teoh},\ and\
  \citenamefont {Hobbs}}]{Manchester_2005_pulsar_catalogue}%
  \BibitemOpen
  \bibfield  {author} {\bibinfo {author} {\bibfnamefont {R.~N.}\ \bibnamefont
  {Manchester}}, \bibinfo {author} {\bibfnamefont {G.~B.}\ \bibnamefont
  {Hobbs}}, \bibinfo {author} {\bibfnamefont {A.}~\bibnamefont {Teoh}}, \ and\
  \bibinfo {author} {\bibfnamefont {M.}~\bibnamefont {Hobbs}},\ }\href
  {\doibase 10.1086/428488} {\bibfield  {journal} {\bibinfo  {journal} {The
  Astronomical Journal}\ }\textbf {\bibinfo {volume} {129}},\ \bibinfo {pages}
  {1993} (\bibinfo {year} {2005})}\BibitemShut {NoStop}%
\bibitem [{\citenamefont {Dexheimer}\ \emph {et~al.}(2017)\citenamefont
  {Dexheimer}, \citenamefont {Franzon}, \citenamefont {Gomes}, \citenamefont
  {Farias},\ and\ \citenamefont
  {Avancini}}]{Dexheimer_2017_field_configurations}%
  \BibitemOpen
  \bibfield  {author} {\bibinfo {author} {\bibfnamefont {V.}~\bibnamefont
  {Dexheimer}}, \bibinfo {author} {\bibfnamefont {B.}~\bibnamefont {Franzon}},
  \bibinfo {author} {\bibfnamefont {R.}~\bibnamefont {Gomes}}, \bibinfo
  {author} {\bibfnamefont {R.}~\bibnamefont {Farias}}, \ and\ \bibinfo {author}
  {\bibfnamefont {S.}~\bibnamefont {Avancini}},\ }\href {\doibase
  https://doi.org/10.1002/asna.201713434} {\bibfield  {journal} {\bibinfo
  {journal} {Astronomische Nachrichten}\ }\textbf {\bibinfo {volume} {338}},\
  \bibinfo {pages} {1052} (\bibinfo {year} {2017})}\BibitemShut {NoStop}%
\bibitem [{\citenamefont {Igoshev}\ \emph {et~al.}(2021)\citenamefont
  {Igoshev}, \citenamefont {Popov},\ and\ \citenamefont
  {Hollerbach}}]{universeMagnetar}%
  \BibitemOpen
  \bibfield  {author} {\bibinfo {author} {\bibfnamefont {A.~P.}\ \bibnamefont
  {Igoshev}}, \bibinfo {author} {\bibfnamefont {S.~B.}\ \bibnamefont {Popov}},
  \ and\ \bibinfo {author} {\bibfnamefont {R.}~\bibnamefont {Hollerbach}},\
  }\href {\doibase 10.3390/universe7090351} {\bibfield  {journal} {\bibinfo
  {journal} {Universe}\ }\textbf {\bibinfo {volume} {7}} (\bibinfo {year}
  {2021}),\ 10.3390/universe7090351}\BibitemShut {NoStop}%
\bibitem [{\citenamefont {Canuto}\ and\ \citenamefont
  {Chiu}(1968)}]{Canuto_chiu_1968}%
  \BibitemOpen
  \bibfield  {author} {\bibinfo {author} {\bibfnamefont {V.}~\bibnamefont
  {Canuto}}\ and\ \bibinfo {author} {\bibfnamefont {H.-Y.}\ \bibnamefont
  {Chiu}},\ }\href {\doibase 10.1103/PhysRev.173.1210} {\bibfield  {journal}
  {\bibinfo  {journal} {Phys. Rev.}\ }\textbf {\bibinfo {volume} {173}},\
  \bibinfo {pages} {1210} (\bibinfo {year} {1968})}\BibitemShut {NoStop}%
\bibitem [{\citenamefont {Tambe}\ \emph {et~al.}(2024)\citenamefont {Tambe},
  \citenamefont {Chatterjee}, \citenamefont {Alford},\ and\ \citenamefont
  {Haber}}]{tambe2024effectmagneticfieldsurca}%
  \BibitemOpen
  \bibfield  {author} {\bibinfo {author} {\bibfnamefont {P.}~\bibnamefont
  {Tambe}}, \bibinfo {author} {\bibfnamefont {D.}~\bibnamefont {Chatterjee}},
  \bibinfo {author} {\bibfnamefont {M.}~\bibnamefont {Alford}}, \ and\ \bibinfo
  {author} {\bibfnamefont {A.}~\bibnamefont {Haber}},\ }\href@noop {} {\
  (\bibinfo {year} {2024})},\ \Eprint {http://arxiv.org/abs/2409.09423}
  {arXiv:2409.09423 [nucl-th]} \BibitemShut {NoStop}%
\bibitem [{\citenamefont {Alford}\ and\ \citenamefont
  {Harris}(2018)}]{AlfordHarrisPRC}%
  \BibitemOpen
  \bibfield  {author} {\bibinfo {author} {\bibfnamefont {M.~G.}\ \bibnamefont
  {Alford}}\ and\ \bibinfo {author} {\bibfnamefont {S.~P.}\ \bibnamefont
  {Harris}},\ }\href {\doibase 10.1103/PhysRevC.98.065806} {\bibfield
  {journal} {\bibinfo  {journal} {Phys. Rev. C}\ }\textbf {\bibinfo {volume}
  {98}},\ \bibinfo {pages} {065806} (\bibinfo {year} {2018})}\BibitemShut
  {NoStop}%
\bibitem [{\citenamefont {Alford}\ \emph {et~al.}(2021)\citenamefont {Alford},
  \citenamefont {Haber}, \citenamefont {Harris},\ and\ \citenamefont
  {Zhang}}]{universeAlfordHaberHarris}%
  \BibitemOpen
  \bibfield  {author} {\bibinfo {author} {\bibfnamefont {M.~G.}\ \bibnamefont
  {Alford}}, \bibinfo {author} {\bibfnamefont {A.}~\bibnamefont {Haber}},
  \bibinfo {author} {\bibfnamefont {S.~P.}\ \bibnamefont {Harris}}, \ and\
  \bibinfo {author} {\bibfnamefont {Z.}~\bibnamefont {Zhang}},\ }\href@noop {}
  {\bibfield  {journal} {\bibinfo  {journal} {Universe}\ }\textbf {\bibinfo
  {volume} {7}} (\bibinfo {year} {2021})}\BibitemShut {NoStop}%
\bibitem [{\citenamefont {Maruyama}\ \emph {et~al.}(2022)\citenamefont
  {Maruyama}, \citenamefont {Balantekin}, \citenamefont {Cheoun}, \citenamefont
  {Kajino}, \citenamefont {Kusakabe},\ and\ \citenamefont
  {Mathews}}]{MARUYAMA2022136813}%
  \BibitemOpen
  \bibfield  {author} {\bibinfo {author} {\bibfnamefont {T.}~\bibnamefont
  {Maruyama}}, \bibinfo {author} {\bibfnamefont {A.~B.}\ \bibnamefont
  {Balantekin}}, \bibinfo {author} {\bibfnamefont {M.-K.}\ \bibnamefont
  {Cheoun}}, \bibinfo {author} {\bibfnamefont {T.}~\bibnamefont {Kajino}},
  \bibinfo {author} {\bibfnamefont {M.}~\bibnamefont {Kusakabe}}, \ and\
  \bibinfo {author} {\bibfnamefont {G.~J.}\ \bibnamefont {Mathews}},\ }\href
  {\doibase https://doi.org/10.1016/j.physletb.2021.136813} {\bibfield
  {journal} {\bibinfo  {journal} {Physics Letters B}\ }\textbf {\bibinfo
  {volume} {824}},\ \bibinfo {pages} {136813} (\bibinfo {year}
  {2022})}\BibitemShut {NoStop}%
\bibitem [{\citenamefont {Duan}\ and\ \citenamefont
  {Qian}(2004)}]{Duan_qian_2004}%
  \BibitemOpen
  \bibfield  {author} {\bibinfo {author} {\bibfnamefont {H.}~\bibnamefont
  {Duan}}\ and\ \bibinfo {author} {\bibfnamefont {Y.-Z.}\ \bibnamefont
  {Qian}},\ }\href {\doibase 10.1103/PhysRevD.69.123004} {\bibfield  {journal}
  {\bibinfo  {journal} {Phys. Rev. D}\ }\textbf {\bibinfo {volume} {69}},\
  \bibinfo {pages} {123004} (\bibinfo {year} {2004})}\BibitemShut {NoStop}%
\bibitem [{\citenamefont {Duan}\ and\ \citenamefont
  {Qian}(2005)}]{duan_qian_2005_appdx_has_integrals}%
  \BibitemOpen
  \bibfield  {author} {\bibinfo {author} {\bibfnamefont {H.}~\bibnamefont
  {Duan}}\ and\ \bibinfo {author} {\bibfnamefont {Y.-Z.}\ \bibnamefont
  {Qian}},\ }\href {\doibase 10.1103/PhysRevD.72.023005} {\bibfield  {journal}
  {\bibinfo  {journal} {Phys. Rev. D}\ }\textbf {\bibinfo {volume} {72}},\
  \bibinfo {pages} {023005} (\bibinfo {year} {2005})}\BibitemShut {NoStop}%
\bibitem [{\citenamefont {Dutra}\ \emph {et~al.}(2014)\citenamefont {Dutra},
  \citenamefont {Louren\ifmmode~\mbox{\c{c}}\else \c{c}\fi{}o}, \citenamefont
  {Avancini}, \citenamefont {Carlson}, \citenamefont {Delfino}, \citenamefont
  {Menezes}, \citenamefont {Provid\^encia}, \citenamefont {Typel},\ and\
  \citenamefont {Stone}}]{PRC_dutra_rmf_review}%
  \BibitemOpen
  \bibfield  {author} {\bibinfo {author} {\bibfnamefont {M.}~\bibnamefont
  {Dutra}}, \bibinfo {author} {\bibfnamefont {O.}~\bibnamefont
  {Louren\ifmmode~\mbox{\c{c}}\else \c{c}\fi{}o}}, \bibinfo {author}
  {\bibfnamefont {S.~S.}\ \bibnamefont {Avancini}}, \bibinfo {author}
  {\bibfnamefont {B.~V.}\ \bibnamefont {Carlson}}, \bibinfo {author}
  {\bibfnamefont {A.}~\bibnamefont {Delfino}}, \bibinfo {author} {\bibfnamefont
  {D.~P.}\ \bibnamefont {Menezes}}, \bibinfo {author} {\bibfnamefont
  {C.}~\bibnamefont {Provid\^encia}}, \bibinfo {author} {\bibfnamefont
  {S.}~\bibnamefont {Typel}}, \ and\ \bibinfo {author} {\bibfnamefont {J.~R.}\
  \bibnamefont {Stone}},\ }\href {\doibase 10.1103/PhysRevC.90.055203}
  {\bibfield  {journal} {\bibinfo  {journal} {Phys. Rev. C}\ }\textbf {\bibinfo
  {volume} {90}},\ \bibinfo {pages} {055203} (\bibinfo {year}
  {2014})}\BibitemShut {NoStop}%
\bibitem [{\citenamefont {Agrawal}\ \emph {et~al.}(2012)\citenamefont
  {Agrawal}, \citenamefont {Sulaksono},\ and\ \citenamefont
  {Reinhard}}]{AGRAWAL20121}%
  \BibitemOpen
  \bibfield  {author} {\bibinfo {author} {\bibfnamefont {B.}~\bibnamefont
  {Agrawal}}, \bibinfo {author} {\bibfnamefont {A.}~\bibnamefont {Sulaksono}},
  \ and\ \bibinfo {author} {\bibfnamefont {P.-G.}\ \bibnamefont {Reinhard}},\
  }\href {\doibase https://doi.org/10.1016/j.nuclphysa.2012.03.004} {\bibfield
  {journal} {\bibinfo  {journal} {Nuclear Physics A}\ }\textbf {\bibinfo
  {volume} {882}},\ \bibinfo {pages} {1} (\bibinfo {year} {2012})}\BibitemShut
  {NoStop}%
\bibitem [{\citenamefont {Cromartie}\ \emph {et~al.}(2019)\citenamefont
  {Cromartie} \emph {et~al.}}]{NANOGrav:2019jur}%
  \BibitemOpen
  \bibfield  {author} {\bibinfo {author} {\bibfnamefont {H.~T.}\ \bibnamefont
  {Cromartie}} \emph {et~al.} (\bibinfo {collaboration} {NANOGrav}),\ }\href
  {\doibase 10.1038/s41550-019-0880-2} {\bibfield  {journal} {\bibinfo
  {journal} {Nature Astron.}\ }\textbf {\bibinfo {volume} {4}},\ \bibinfo
  {pages} {72} (\bibinfo {year} {2019})},\ \Eprint
  {http://arxiv.org/abs/1904.06759} {arXiv:1904.06759 [astro-ph.HE]}
  \BibitemShut {NoStop}%
\bibitem [{\citenamefont {Antoniadis}\ \emph {et~al.}(2013)\citenamefont
  {Antoniadis} \emph {et~al.}}]{Antoniadis:2013pzd}%
  \BibitemOpen
  \bibfield  {author} {\bibinfo {author} {\bibfnamefont {J.}~\bibnamefont
  {Antoniadis}} \emph {et~al.},\ }\href {\doibase 10.1126/science.1233232}
  {\bibfield  {journal} {\bibinfo  {journal} {Science}\ }\textbf {\bibinfo
  {volume} {340}},\ \bibinfo {pages} {6131} (\bibinfo {year} {2013})},\ \Eprint
  {http://arxiv.org/abs/1304.6875} {arXiv:1304.6875 [astro-ph.HE]} \BibitemShut
  {NoStop}%
\bibitem [{\citenamefont {Demorest}\ \emph {et~al.}(2010)\citenamefont
  {Demorest}, \citenamefont {Pennucci}, \citenamefont {Ransom}, \citenamefont
  {Roberts},\ and\ \citenamefont {Hessels}}]{Demorest:2010bx}%
  \BibitemOpen
  \bibfield  {author} {\bibinfo {author} {\bibfnamefont {P.}~\bibnamefont
  {Demorest}}, \bibinfo {author} {\bibfnamefont {T.}~\bibnamefont {Pennucci}},
  \bibinfo {author} {\bibfnamefont {S.}~\bibnamefont {Ransom}}, \bibinfo
  {author} {\bibfnamefont {M.}~\bibnamefont {Roberts}}, \ and\ \bibinfo
  {author} {\bibfnamefont {J.}~\bibnamefont {Hessels}},\ }\href {\doibase
  10.1038/nature09466} {\bibfield  {journal} {\bibinfo  {journal} {Nature}\
  }\textbf {\bibinfo {volume} {467}},\ \bibinfo {pages} {1081} (\bibinfo {year}
  {2010})},\ \Eprint {http://arxiv.org/abs/1010.5788} {arXiv:1010.5788
  [astro-ph.HE]} \BibitemShut {NoStop}%
\bibitem [{\citenamefont {Abbott}\ \emph {et~al.}(2018)\citenamefont {Abbott}
  \emph {et~al.}}]{LIGOScientific:2018cki}%
  \BibitemOpen
  \bibfield  {author} {\bibinfo {author} {\bibfnamefont {B.~P.}\ \bibnamefont
  {Abbott}} \emph {et~al.} (\bibinfo {collaboration} {LIGO Scientific,
  Virgo}),\ }\href {\doibase 10.1103/PhysRevLett.121.161101} {\bibfield
  {journal} {\bibinfo  {journal} {Phys. Rev. Lett.}\ }\textbf {\bibinfo
  {volume} {121}},\ \bibinfo {pages} {161101} (\bibinfo {year} {2018})},\
  \Eprint {http://arxiv.org/abs/1805.11581} {arXiv:1805.11581 [gr-qc]}
  \BibitemShut {NoStop}%
\bibitem [{\citenamefont {De}\ \emph {et~al.}(2018)\citenamefont {De},
  \citenamefont {Finstad}, \citenamefont {Lattimer}, \citenamefont {Brown},
  \citenamefont {Berger},\ and\ \citenamefont {Biwer}}]{De:2018uhw}%
  \BibitemOpen
  \bibfield  {author} {\bibinfo {author} {\bibfnamefont {S.}~\bibnamefont
  {De}}, \bibinfo {author} {\bibfnamefont {D.}~\bibnamefont {Finstad}},
  \bibinfo {author} {\bibfnamefont {J.~M.}\ \bibnamefont {Lattimer}}, \bibinfo
  {author} {\bibfnamefont {D.~A.}\ \bibnamefont {Brown}}, \bibinfo {author}
  {\bibfnamefont {E.}~\bibnamefont {Berger}}, \ and\ \bibinfo {author}
  {\bibfnamefont {C.~M.}\ \bibnamefont {Biwer}},\ }\href {\doibase
  10.1103/PhysRevLett.121.091102} {\bibfield  {journal} {\bibinfo  {journal}
  {Phys. Rev. Lett.}\ }\textbf {\bibinfo {volume} {121}},\ \bibinfo {pages}
  {091102} (\bibinfo {year} {2018})},\ \bibinfo {note} {[Erratum:
  Phys.Rev.Lett. 121, 259902 (2018)]},\ \Eprint
  {http://arxiv.org/abs/1804.08583} {arXiv:1804.08583 [astro-ph.HE]}
  \BibitemShut {NoStop}%
\bibitem [{\citenamefont {Capano}\ \emph {et~al.}(2020)\citenamefont {Capano},
  \citenamefont {Tews}, \citenamefont {Brown}, \citenamefont {Margalit},
  \citenamefont {De}, \citenamefont {Kumar}, \citenamefont {Brown},
  \citenamefont {Krishnan},\ and\ \citenamefont {Reddy}}]{Capano:2019eae}%
  \BibitemOpen
  \bibfield  {author} {\bibinfo {author} {\bibfnamefont {C.~D.}\ \bibnamefont
  {Capano}}, \bibinfo {author} {\bibfnamefont {I.}~\bibnamefont {Tews}},
  \bibinfo {author} {\bibfnamefont {S.~M.}\ \bibnamefont {Brown}}, \bibinfo
  {author} {\bibfnamefont {B.}~\bibnamefont {Margalit}}, \bibinfo {author}
  {\bibfnamefont {S.}~\bibnamefont {De}}, \bibinfo {author} {\bibfnamefont
  {S.}~\bibnamefont {Kumar}}, \bibinfo {author} {\bibfnamefont {D.~A.}\
  \bibnamefont {Brown}}, \bibinfo {author} {\bibfnamefont {B.}~\bibnamefont
  {Krishnan}}, \ and\ \bibinfo {author} {\bibfnamefont {S.}~\bibnamefont
  {Reddy}},\ }\href {\doibase 10.1038/s41550-020-1014-6} {\bibfield  {journal}
  {\bibinfo  {journal} {Nature Astron.}\ }\textbf {\bibinfo {volume} {4}},\
  \bibinfo {pages} {625} (\bibinfo {year} {2020})},\ \Eprint
  {http://arxiv.org/abs/1908.10352} {arXiv:1908.10352 [astro-ph.HE]}
  \BibitemShut {NoStop}%
\bibitem [{\citenamefont {Miller}\ \emph {et~al.}(2019)\citenamefont {Miller}
  \emph {et~al.}}]{Miller:2019cac}%
  \BibitemOpen
  \bibfield  {author} {\bibinfo {author} {\bibfnamefont {M.~C.}\ \bibnamefont
  {Miller}} \emph {et~al.},\ }\href {\doibase 10.3847/2041-8213/ab50c5}
  {\bibfield  {journal} {\bibinfo  {journal} {Astrophys. J. Lett.}\ }\textbf
  {\bibinfo {volume} {887}},\ \bibinfo {pages} {L24} (\bibinfo {year}
  {2019})},\ \Eprint {http://arxiv.org/abs/1912.05705} {arXiv:1912.05705
  [astro-ph.HE]} \BibitemShut {NoStop}%
\bibitem [{\citenamefont {Riley}\ \emph {et~al.}(2019)\citenamefont {Riley}
  \emph {et~al.}}]{Riley:2019yda}%
  \BibitemOpen
  \bibfield  {author} {\bibinfo {author} {\bibfnamefont {T.~E.}\ \bibnamefont
  {Riley}} \emph {et~al.},\ }\href {\doibase 10.3847/2041-8213/ab481c}
  {\bibfield  {journal} {\bibinfo  {journal} {Astrophys. J. Lett.}\ }\textbf
  {\bibinfo {volume} {887}},\ \bibinfo {pages} {L21} (\bibinfo {year}
  {2019})},\ \Eprint {http://arxiv.org/abs/1912.05702} {arXiv:1912.05702
  [astro-ph.HE]} \BibitemShut {NoStop}%
\bibitem [{\citenamefont {Steinmetz}\ \emph {et~al.}(2019)\citenamefont
  {Steinmetz}, \citenamefont {Formanek},\ and\ \citenamefont
  {Rafelski}}]{Steinmetz2019KGPvsDP}%
  \BibitemOpen
  \bibfield  {author} {\bibinfo {author} {\bibfnamefont {A.}~\bibnamefont
  {Steinmetz}}, \bibinfo {author} {\bibfnamefont {M.}~\bibnamefont {Formanek}},
  \ and\ \bibinfo {author} {\bibfnamefont {J.}~\bibnamefont {Rafelski}},\
  }\href {\doibase 10.1140/epja/i2019-12715-5} {\bibfield  {journal} {\bibinfo
  {journal} {The European Physical Journal A}\ }\textbf {\bibinfo {volume}
  {55}},\ \bibinfo {pages} {40} (\bibinfo {year} {2019})}\BibitemShut {NoStop}%
\bibitem [{\citenamefont {Broderick}\ \emph {et~al.}(2000)\citenamefont
  {Broderick}, \citenamefont {Prakash},\ and\ \citenamefont
  {Lattimer}}]{Broderick_2000PLmageos}%
  \BibitemOpen
  \bibfield  {author} {\bibinfo {author} {\bibfnamefont {A.}~\bibnamefont
  {Broderick}}, \bibinfo {author} {\bibfnamefont {M.}~\bibnamefont {Prakash}},
  \ and\ \bibinfo {author} {\bibfnamefont {J.~M.}\ \bibnamefont {Lattimer}},\
  }\href {\doibase 10.1086/309010} {\bibfield  {journal} {\bibinfo  {journal}
  {The Astrophysical Journal}\ }\textbf {\bibinfo {volume} {537}},\ \bibinfo
  {pages} {351–367} (\bibinfo {year} {2000})}\BibitemShut {NoStop}%
\bibitem [{\citenamefont {Gradshteyn}\ and\ \citenamefont
  {Ryzhik}(1980)}]{integralsseriesproducts}%
  \BibitemOpen
  \bibfield  {author} {\bibinfo {author} {\bibfnamefont {I.~S.}\ \bibnamefont
  {Gradshteyn}}\ and\ \bibinfo {author} {\bibfnamefont {I.~M.}\ \bibnamefont
  {Ryzhik}},\ }\href@noop {} {\emph {\bibinfo {title} {Table of Integrals,
  Series, and Products}}}\ (\bibinfo  {publisher} {Academic Press},\ \bibinfo
  {address} {New York},\ \bibinfo {year} {1980})\BibitemShut {NoStop}%
\bibitem [{\citenamefont {Bhattacharya}(2004)}]{bhattacharya_thesis}%
  \BibitemOpen
  \bibfield  {author} {\bibinfo {author} {\bibfnamefont {K.}~\bibnamefont
  {Bhattacharya}},\ }\emph {\bibinfo {title} {Elementary Particle Interactions
  In A Background Magnetic Field}},\ \href@noop {} {Ph.D. thesis},\ \bibinfo
  {school} {Jadavpur University} (\bibinfo {year} {2004}),\ \Eprint
  {http://arxiv.org/abs/hep-ph/0407099} {arXiv:hep-ph/0407099 [hep-ph]}
  \BibitemShut {NoStop}%
\bibitem [{\citenamefont {Shapiro}\ and\ \citenamefont
  {Teukolsky}(1983)}]{ShapiroTeukolsky1983}%
  \BibitemOpen
  \bibfield  {author} {\bibinfo {author} {\bibfnamefont {S.~L.}\ \bibnamefont
  {Shapiro}}\ and\ \bibinfo {author} {\bibfnamefont {S.~A.}\ \bibnamefont
  {Teukolsky}},\ }\href@noop {} {\emph {\bibinfo {title} {Black Holes, White
  Dwarfs, and Neutron Stars: The Physics of Compact Objects}}}\ (\bibinfo
  {publisher} {Wiley},\ \bibinfo {address} {New York},\ \bibinfo {year}
  {1983})\BibitemShut {NoStop}%
\bibitem [{\citenamefont {Most}\ \emph {et~al.}(2021)\citenamefont {Most},
  \citenamefont {Harris}, \citenamefont {Plumberg}, \citenamefont {Alford},
  \citenamefont {Noronha}, \citenamefont {Noronha-Hostler}, \citenamefont
  {Pretorius}, \citenamefont {Witek},\ and\ \citenamefont
  {Yunes}}]{Harris_urcaviscosity}%
  \BibitemOpen
  \bibfield  {author} {\bibinfo {author} {\bibfnamefont {E.~R.}\ \bibnamefont
  {Most}}, \bibinfo {author} {\bibfnamefont {S.~P.}\ \bibnamefont {Harris}},
  \bibinfo {author} {\bibfnamefont {C.}~\bibnamefont {Plumberg}}, \bibinfo
  {author} {\bibfnamefont {M.~G.}\ \bibnamefont {Alford}}, \bibinfo {author}
  {\bibfnamefont {J.}~\bibnamefont {Noronha}}, \bibinfo {author} {\bibfnamefont
  {J.}~\bibnamefont {Noronha-Hostler}}, \bibinfo {author} {\bibfnamefont
  {F.}~\bibnamefont {Pretorius}}, \bibinfo {author} {\bibfnamefont
  {H.}~\bibnamefont {Witek}}, \ and\ \bibinfo {author} {\bibfnamefont
  {N.}~\bibnamefont {Yunes}},\ }\href {\doibase 10.1093/mnras/stab2793}
  {\bibfield  {journal} {\bibinfo  {journal} {Monthly Notices of the Royal
  Astronomical Society}\ }\textbf {\bibinfo {volume} {509}},\ \bibinfo {pages}
  {1096} (\bibinfo {year} {2021})}\BibitemShut {NoStop}%
\bibitem [{\citenamefont {Endrizzi}\ \emph {et~al.}(2020)\citenamefont
  {Endrizzi}, \citenamefont {Perego}, \citenamefont {Fabbri}, \citenamefont
  {Branca}, \citenamefont {Radice}, \citenamefont {Bernuzzi}, \citenamefont
  {Giacomazzo}, \citenamefont {Pederiva},\ and\ \citenamefont
  {Lovato}}]{Endrizzi_2020}%
  \BibitemOpen
  \bibfield  {author} {\bibinfo {author} {\bibfnamefont {A.}~\bibnamefont
  {Endrizzi}}, \bibinfo {author} {\bibfnamefont {A.}~\bibnamefont {Perego}},
  \bibinfo {author} {\bibfnamefont {F.~M.}\ \bibnamefont {Fabbri}}, \bibinfo
  {author} {\bibfnamefont {L.}~\bibnamefont {Branca}}, \bibinfo {author}
  {\bibfnamefont {D.}~\bibnamefont {Radice}}, \bibinfo {author} {\bibfnamefont
  {S.}~\bibnamefont {Bernuzzi}}, \bibinfo {author} {\bibfnamefont
  {B.}~\bibnamefont {Giacomazzo}}, \bibinfo {author} {\bibfnamefont
  {F.}~\bibnamefont {Pederiva}}, \ and\ \bibinfo {author} {\bibfnamefont
  {A.}~\bibnamefont {Lovato}},\ }\href@noop {} {\bibfield  {journal} {\bibinfo
  {journal} {The European Physical Journal A}\ }\textbf {\bibinfo {volume}
  {56}} (\bibinfo {year} {2020})}\BibitemShut {NoStop}%
\bibitem [{\citenamefont {Kiuchi}\ \emph {et~al.}(2015)\citenamefont {Kiuchi},
  \citenamefont {Cerd\'a-Dur\'an}, \citenamefont {Kyutoku}, \citenamefont
  {Sekiguchi},\ and\ \citenamefont {Shibata}}]{PRD_mergerfields}%
  \BibitemOpen
  \bibfield  {author} {\bibinfo {author} {\bibfnamefont {K.}~\bibnamefont
  {Kiuchi}}, \bibinfo {author} {\bibfnamefont {P.}~\bibnamefont
  {Cerd\'a-Dur\'an}}, \bibinfo {author} {\bibfnamefont {K.}~\bibnamefont
  {Kyutoku}}, \bibinfo {author} {\bibfnamefont {Y.}~\bibnamefont {Sekiguchi}},
  \ and\ \bibinfo {author} {\bibfnamefont {M.}~\bibnamefont {Shibata}},\ }\href
  {\doibase 10.1103/PhysRevD.92.124034} {\bibfield  {journal} {\bibinfo
  {journal} {Phys. Rev. D}\ }\textbf {\bibinfo {volume} {92}},\ \bibinfo
  {pages} {124034} (\bibinfo {year} {2015})}\BibitemShut {NoStop}%
\bibitem [{\citenamefont {Price}\ and\ \citenamefont
  {Rosswog}(2006)}]{Science_PriceRosswog}%
  \BibitemOpen
  \bibfield  {author} {\bibinfo {author} {\bibfnamefont {D.~J.}\ \bibnamefont
  {Price}}\ and\ \bibinfo {author} {\bibfnamefont {S.}~\bibnamefont
  {Rosswog}},\ }\href {\doibase 10.1126/science.1125201} {\bibfield  {journal}
  {\bibinfo  {journal} {Science}\ }\textbf {\bibinfo {volume} {312}},\ \bibinfo
  {pages} {719} (\bibinfo {year} {2006})}\BibitemShut {NoStop}%
\bibitem [{\citenamefont {Kumamoto}\ and\ \citenamefont {Welch}(2024)}]{nslq}%
  \BibitemOpen
  \bibfield  {author} {\bibinfo {author} {\bibfnamefont {M.}~\bibnamefont
  {Kumamoto}}\ and\ \bibinfo {author} {\bibfnamefont {C.}~\bibnamefont
  {Welch}},\ }\href {https://github.com/clwelch03/NS-landau-quantization}
  {\enquote {\bibinfo {title} {{NS-landau-quantization}},}\ } (\bibinfo {year}
  {{2024}}),\ \bibinfo {note} {github repository}\BibitemShut {NoStop}%
\end{thebibliography}%
\end{document}